\documentclass[twocolumn,english]{revtex4-2}
\usepackage[T1]{fontenc}
\usepackage{color}
\usepackage{babel}
\usepackage{mathtools}
\usepackage{bm}
\usepackage{graphicx}
\usepackage{esint}
\usepackage{amsfonts}
\usepackage{graphicx}
\usepackage{float}
\usepackage{subfigure}
\usepackage{amsthm}
\usepackage{amssymb}
\usepackage[ruled, longend]{algorithm2e}
\usepackage{multirow}
\usepackage{physics}

\usepackage[usenames,divipsnames,svgnames,table]{xcolor}
\usepackage[unicode=true,pdfusetitle,
bookmarks=true,bookmarksnumbered=false,bookmarksopen=false,
breaklinks=true,pdfborder={0 0 0},backref=false,colorlinks=true]{hyperref}
\hypersetup{linkcolor=NavyBlue,urlcolor=NavyBlue,citecolor=NavyBlue}
\usepackage{breakurl}

\usepackage{orcidlink}

\begin{document}

\title{Quantum speed limit for complex dynamics}

\author{Mao Zhang}
\affiliation{National Precise Gravity Measurement Facility, MOE Key
Laboratory of Fundamental Physical Quantities Measurement, School of Physics,
Huazhong University of Science and Technology, Wuhan 430074, China}

\author{Huai-Ming Yu}
\affiliation{National Precise Gravity Measurement Facility, MOE Key
Laboratory of Fundamental Physical Quantities Measurement, School of Physics,
Huazhong University of Science and Technology, Wuhan 430074, China}

\author{Jing Liu\orcidlink{0000-0001-9944-4493}}
\email{liujingphys@hust.edu.cn}
\affiliation{National Precise Gravity Measurement Facility, MOE Key
Laboratory of Fundamental Physical Quantities Measurement, School of Physics,
Huazhong University of Science and Technology, Wuhan 430074, China}

\begin{abstract}
Quantum speed limit focuses on the minimum time scale for a fixed mission and hence is important in quantum information 
where fast dynamics is usually beneficial. Most existing tools for the depiction of quantum speed limit are the 
lower-bound-type tools, which are in fact difficult to reveal the true minimum time, especially for many-body systems 
or complex dynamics. Therefore, the evaluation of this true minimum time in these scenarios is still an unsolved problem. 
Hereby we propose a three-step (classification-regression-calibration) methodology based on machine learning to evaluate 
the true minimum time in complex dynamics. Moreover, the analytical expression of the true minimum time is also provided 
for the time-dependent Hamiltonians with time-independent eigenstates. 
\end{abstract}

\maketitle

Quantum speed limit (QSL) is a fundamental topic in quantum mechanics focusing on the characterization of minimum 
time for quantum states to fulfill certain known targets, such as rotating a state to its orthogonal states, or some 
angles quantified by certain metrics. In principle, the target could be chosen flexibly due to the problem of interest.  
In the year of 1945, Mandelstam and Tamm provided the first lower bound for this minimum time based on the uncertainty 
relation~\cite{Mandelstam1945}. In 1996 Braunstein \emph{et al.} extended the lower bound to time-dependent Hamiltonians 
utilizing the generalized uncertainty relation~\cite{Braunstein1996} where the time-average variance was applied. In 1998, 
Margolus and Levitin~\cite{Margolus1998} provided another bound based on the mean energy. After these pioneer works, the 
topic of QSL entered a period of rapid development in the next 20 years, especially in 2010s~\cite{Giovannetti2003,
Giovannetti2004,Levitin2009,Caneva2009,Taddei2013,Campo2013,Deffner2013,Hegerfeldt2013,Liu2015,Marvian2015,Sun2015,
Pires2016,Deffner2017,Chenu2017,Beau2017a,Beau2017b,Funo2017,Campbell2017,Cai2017,Campaioli2018,Campaioli2019,
Shanahan2018,Okuyama2018,Wu2018,Bukov2019,Girolami2019,Sun2019,Hu2020,Shao2020,Sun2021,Becker2021,Liu2021,
Ness2021,Campo2021,Ness2022,Pintos2022,Liu2020,Wu2022,Mondal2016}.

Most existing tools in QSL belong to the lower-bound-type (LBT) tools. The advantage of this type of tools is that they are 
easy to compute, especially in numerical aspects. However, the disadvantage of them are also significant. On one hand, most 
LBT tools are dependent on the initial states. This dependence would cause a problem that even the initial state cannot 
actually fulfill the given target, the LBT tools would still provide finite results, which is reasonable in mathematics since 
any finite value is a legitimate lower bound of infinity. However, it also indicates that from these tools one cannot acquire 
the information whether a state is capable to fulfill the target. For example, consider a qubit Hamiltonian $\omega\sigma_z/2$ 
with $\sigma_{z(x)}$ the Pauli Z (X) matrix and $\omega$ the energy gap. For this Hamiltonian, the Mandelstam-Tamm and 
Margolus-Levitin bounds for the state with the density matrix $\openone/2+\sigma_x/4+\sqrt{3}\sigma_z/4$ are $2\pi/\omega$ 
and $2\pi/(\sqrt{3}\omega)$. Here $\openone$ is the identity matrix. However, in fact this state cannot fulfill the target $\pi/2$ 
at all since the maximum angle it  can rotate under the given Hamiltonian is only $\pi/3$~\cite{Shao2020}. Hence, without the 
information whether the target can be fulfilled, the conclusions based on the lower-bound-type tools might be suboptimal 
since the results are actually unphysical for the states unable to reach the target. 

On the other hand, in the case that the Hamiltonians are time-dependent, the LBT tools are usually functions of 
time~\cite{Deffner2013,Sun2015,Marvian2015,Funo2017,Shanahan2018,Campo2013,Wu2018,Campaioli2018,Campaioli2019,
Sun2019,Sun2021,Pires2016}. As a matter of fact, these formal time-dependent lower bounds are difficult to reveal both the 
true minimum time and true physics behind it. As clarified in Ref.~\cite{Shao2020}, in noncontrolled scenarios the true 
minimum time for a fixed state to fulfill a given target is only a fixed time point, and the results of LBT tools have to go across 
this time point due to their time dependence. During the time before this time point, the finite results of the LBT tools cannot 
reveal the fact that this state is actually uncapable to reach the target in the time regime. And during the time after this time 
point, the results of LBT tools have to be no larger than this point since they are its lower bounds, which indicates that in this 
time regime the attainability of the LBT tools is lousy. These disadvantages of LBT tools could be further magnified with the 
growth of system dimension or the complexity of dynamics. Hence, locating the true minimum time for the fulfillness of a given 
target in many-body systems and complex dynamics is still an important yet unsolved problem. Finding this minimum time or at 
least providing efficient methodologies to search it is thus the major motivation of this paper. 

\vspace{2ex}
\textbf{\large{Results and discussion}}

\vspace{0.5ex}
\textbf{Operational definition of the quantum speed limit}

The target in QSL could be quantified via different tools, such as the Bures metric or various types of fidelity~\cite{Taddei2013,
Deffner2013,Sun2015,Marvian2015,Funo2017,Shanahan2018}, relative purity~\cite{Campo2013,Wu2018,Campaioli2018}, Bloch 
angle~\cite{Campaioli2018,Campaioli2019,Shao2020,Liu2021}, gauge invariant distances~\cite{Sun2019,Sun2021}, and 
Wigner-Yanase information~\cite{Pires2016}. Different tools usually lead to different mathematical bounds or methods for the 
description of QSL, and a general and unified methodology that fits all tools is still in lack. Recently, an operational definition of 
the quantum speed limit (OQSL) was proposed~\cite{Shao2020} based on the Bloch angle, which is capable to be extended to a 
general tool due to the fact that it is intrinsically a methodology, rather than a concept. Denote $\rho$ as the density matrix of a 
quantum state, $\Phi$ as any type of metric or tool to quantify the target and $\Phi_{\mathrm{tar}}$ as the corresponding target 
value, then the reachable state set can be defined as $\mathcal{S}:=\{\rho\,|\,\Phi(t,\rho)=\Phi_{\mathrm{tar}},\exists t\}$, which 
is the set of states that can fulfill the target. Moreover, it is possible that in some cases not all states in the state space, but the 
states in a subset $\mathcal{Q}$,  are concerned. In this case, $\mathcal{S}$ can be further expressed by 
$\mathcal{S}:=\{\rho\,|\,\rho\in \mathcal{Q} ~\&~\Phi(t,\rho)=\Phi_{\mathrm{tar}},\exists t\}$. Utilizing $\mathcal{S}$, the OQSL 
(denoted by $\tau$) can be defined by 
\begin{eqnarray}
\tau &:=& \min_{\rho\in\mathcal{S}} t \nonumber \\
& & \mathrm{subject~to}~\Phi(t,\rho)=\Phi_{\mathrm{tar}}.
\end{eqnarray}

The Bloch vector is one of the most famous geometric representations for the quantum state and has been widely applied in many 
fields of quantum physics, such as the quantum computation~\cite{Lanyon2013} and quantum control~\cite{Kim2014}.  In the Bloch 
representation, the density matrix can be expressed by $\rho=\frac{1}{N}\big(\openone+\sqrt{\frac{1}{2}N(N-1)}\vec{r}
\cdot\vec{\lambda}\big)$, where $N$ is the dimension of $\rho$, $\vec{\lambda}$ is the vector of SU($N$) generators, $\vec{r}$ is 
the Bloch vector satisfying $|\vec{r}|\leq1$, and $\openone$ is the identity matrix. The Bloch angle $\theta$ between $\vec{r}$ and its 
evolved vector $\vec{r}(t)$ is $\theta(t,\vec{r}):=\arccos\big(\frac{\vec{r}\cdot\vec{r}(t)}{|\vec{r}||\vec{r}(t)|}\big)\in(0,\pi]$. 
Denote $\Theta$ as the fixed target, then the reachable state set can be rewritten into $\mathcal{S}=\{\vec{r}\,|\,\vec{r}\in\mathcal{Q}
~\&~\theta(t,\vec{r})=\Theta,\exists t\}$, and the OQSL reads $\tau=\min_{\vec{r}\in\mathcal{S}}\,t$, subjecting to the constraint 
$\theta(t,\vec{r})=\Theta$. 

In the perspective of OQSL, when two tools to quantify the target has a one-to-one correspondence, for example the angle of relative 
purity $\arccos\big(\frac{\mathrm{Tr}(\rho\rho(t))}{\mathrm{Tr}(\rho^2)}\big)$ and Bloch angle (calculation details are in the 
Supplementary Information), then the reachable state sets for these tools are exactly the same, which means the results of OQSL would 
also be equivalent. This equivalence reveals an important fact that a physical target can be mathematically quantified by different tools, 
yet the true minimum time to fulfill the physical target should not be affected by this quantification process since it is not physical. 

The OQSL is closely related to the quantum brachistochrone problem~\cite{Carlini2006,Carlini2007}, which focuses on searching the 
minimum time for a given initial state to a fixed target state or the realization of a target gate. In the language of OQSL, instead of a given 
initial state, we can study the minimum time for a set of initial states, i.e., the aforementioned set $\mathcal{Q}$, to reach a target state 
$\rho_{\mathrm{tar}}$ under a given Hamiltonian. In this problem $\mathcal{S}$ can be expressed by $\mathcal{S}=\{\rho|\rho\in\mathcal{Q}
~\&~e^{\mathcal{L}}(\rho)=\rho_{\mathrm{tar}}, \exists t\}$ where $\mathcal{L}$ is a superoperator satisfying $\partial_t 
\rho_t=\mathcal{L}(\rho_t)$ with $\rho_t$ the evolved state of $\rho$. Furthermore, the OQSL can be expressed by 
\begin{eqnarray}
\tau &:=& \min_{\rho\in\mathcal{S}} t \nonumber \\
& & \mathrm{subject~to}~e^{\mathcal{L}}(\rho)=\rho_{\mathrm{tar}}.
\end{eqnarray}
Notice that if $\rho_{\mathrm{tar}}\in \mathcal{Q}$, the optimal state 
in $\mathcal{Q}$ to reach $\rho_{\mathrm{tar}}$ must be $\rho_{\mathrm{tar}}$ itself for any Hamiltonian and the corresponding time is 
nothing but zero, which means this is a trivial case. Therefore, $\rho_{\mathrm{tar}}\notin \mathcal{Q}$ should be satisfied to make sure 
the problem is nontrivial. Here we still take the qubit Hamiltonian $\omega\sigma_z/2$ as a simple demonstration. The target state is 
assumed to be $(\ket{0}-\ket{1})/\sqrt{2}$ with $\ket{0}$ ($\ket{1}$) the eigenstate of $\sigma_z$ corresponding to the eigenvalue 
$1$ ($-1$).  $\mathcal{Q}=\{\rho|\mathrm{Tr}(\rho\sigma_x)\geq 0\}$.  Utilizing the spherical coordinates of the Bloch vector 
$\vec{r}=\eta (\sin\alpha\cos\varphi,\sin\alpha\sin\varphi,\cos\alpha)^{\mathrm{T}}$, $\mathcal{S}$ in this example reads 
$\left\{\vec{r}\,|\,\eta=1,\alpha=\pi/2,\varphi\in[0,\pi/2]\cup[3\pi/2,2\pi)\right\}$, and the OQSL $\tau=\pi/(2\omega)$. This minimum 
time can be attained by the state $(\ket{0}+i\ket{1})\sqrt{2}$. Calculation details can be found in the Supplementary Information. 

Compared to lower-bound-type QSLs, the advantages of OQSL are that it can reveal the information that whether a state can fulfill 
the target, and it is always attainable~\cite{Shao2020}. In the case of complex dynamics, these advantages come at a price of 
high computational complexity, which is not only due to the optimization in the definition, but also the preliminary assumption 
that $\mathcal{S}$ is known. For example, in the analytical calculation of the OQSL, the search of $\mathcal{S}$ is the first step and 
usually finished by finding the condition of $\rho$ when the equation $\Phi\left(t,\rho\right)=\Phi_{\mathrm{tar}}$ has a finite solution 
$t$. Then the evolution time to fulfill the target is calculated and optimized under this condition to further obtain the OQSL. In this 
case, the calculation of $\mathcal{S}$ and the optimization of time are performed separably and thus their contributions to the 
computational complexity are different. In the numerical evaluation of OQSL, the contributions of these two processes are the same when 
the brute-force search is applied since the search of $\mathcal{S}$ in this method is based on the rigorous dynamics of each state. 
When $\mathcal{S}$ is obtained, the corresponding time to fulfill the target for each state is also obtained. Hence, the computational 
complexity in this case is basically contributed by the search of $\mathcal{S}$. However, it is obvious that the brute-force search is not 
always feasible in practice, especially when the dynamics is complex or the system size is large, which is actually a non-negligible scenario 
in the study of QSL~\cite{Liu2015,Chenu2017,Beau2017a,Bukov2019}. Hence, finding methods for the evaluation of OQSL that are 
friendly to the complex dynamics or large-size systems is critical, and thus the major motivation of this paper. 

\vspace{2ex}
\textbf{The time-dependent Hamiltonians with time-independent eigenstates}

In many cases, the complexity of dynamics comes from the time dependency of the Hamiltonian. The OQSL for a general
time-dependent Hamiltonian is difficult to obtain analytically. However, for the time-dependent Hamiltonians with
time-independent eigenstates, the OQSL can be obtained analytically when taking the Bloch angle as the quantification of target. 
In the energy space, these Hamiltonians can be expressed by $H(t)=\sum_i E_i(t)\ket{E_i}\bra{E_i}$, where the eigenstate $\ket{E_i}$ 
is time-independent for any $i$ and the eigenvalue $E_i(t)$ depends on time. Many well-known models in quantum mechanics fit this 
scenario, such as the one-dimensional Ising model with a time-varying longitudinal field, the resonant Jaynes-Cummings model with 
time-dependent coupling~\cite{Joshi1993,Lawande1994,Du2022}, and the semiclassical qubit-field model in the strong coupling 
regime~\cite{Scully1997}. For such Hamiltonians, we present the following theorem.

%============================================ Theorem ============================================
\textbf{Theorem.} For a $N$-dimensional time-dependent Hamiltonian whose eigenstates are all
time-independent, the OQSL $\tau$ satisfies the equation
\begin{equation}
\int_0^\tau \left[E_{\max}(t)-E_{\min}(t)\right]\mathrm{d}t=\Theta,
\label{eq:OQSL}
\end{equation}
where $E_{\max}(t)$ and $E_{\min}(t)$ are the maximum and minimum energies of the Hamiltonian at
time $t$. Further denoting the $p$-dimensional set $\{\ket{E_{\min}}\}$ and $q$-dimensional set
$\{\ket{E_{\max}}\}$ as the sets of eigenstates with respect to $E_{\min}(t)$ and $E_{\max}(t)$,
the optimal states to reach the OQSL are
\begin{equation*}
\sum_{i}\frac{1}{N}\ket{E_i}\bra{E_i}+\!\!\!\sum_{\ket{E_k}\in\{\ket{E_{\min}}\},\atop
\ket{E_l}\in\{\ket{E_{\max}}\}}\!\!\!\xi_{kl}\ket{E_k}\bra{E_l}
+\xi_{kl}^{*}\ket{E_l}\bra{E_k},
\end{equation*}
where the matrix $\xi$ (with $kl$th entry $\xi_{kl}$) satisfies $N^2\xi^{\dagger}\xi\leq\openone_{q}$
with $\openone_{q}$ the $q$-dimensional identity matrix.
%================================================================================================

The proof is given in the Supplementary Information. As a matter of fact, this theorem covers Theorem 1 in 
Ref.~\cite{Shao2020} due to the fact that Eq.~(\ref{eq:OQSL}) reduces to $\tau=\Theta/(E_{\max}-E_{\min})$ when 
the eigenvalues are time-independent. As a simple demonstration, consider the Hamiltonian $H(t)=f(t)\sigma_z$ with 
$f(t)$ a time-dependent function. It is obvious that the eigenstates of this Hamiltonian are independent of time. Hence 
the corresponding OQSL is given in the theorem above. In the case that $|\int_0^t f(t_1)\mathrm{d}t_1|$ is upper 
bounded by $c_f$, $\mathcal{S}$ is fully determined by the value of $c_f$, which leads to the following corollary.

%============================================ Theorem ============================================
\textbf{Corollary.} For the Hamiltonian $H(t)=f(t)\sigma_z$ where $f(t)$ satisfies
$|\int_0^t f(t_1)\mathrm{d}t_1|\leq c_f$, no state can fulfill the target $\Theta$ 
if $c_f<\Theta/2$.
%================================================================================================

In the case that $c_f\geq \Theta/2$, $\mathcal{S}$ is symmetric about $z$ axis in the Bloch sphere, similar to the time-independent 
Hamiltonian $\omega\sigma_z/2$~\cite{Shao2020}. This is due to the fact that in this case the dynamics of all states in the 
Bloch sphere are the precessions about $z$ axis, and thus it obeys the rotational symmetry about $z$ axis. Therefore, 
$\mathcal{S}$ can be fully expressed by the angle between the Bloch vector and $z$ axis (denoted by $\alpha$). More 
specifically to say, when $c_f\in [\Theta/2,\pi/2]$, $\mathcal{S}=\{\vec{r}\,|\alpha\!\in\![\alpha_f,\pi-\alpha_f]\}$ with 
$\alpha_f=\arcsin\left(\frac{\sin(\Theta/2)}{\sin c_f}\right)$, and $\mathcal{S}=\{\vec{r}\,|\alpha\!\in\![\Theta/2,\pi-\Theta/2]\}$ 
when $c_f\!>\!\pi/2$. Furthermore, the OQSL satisfies $\int^{\tau}_0 |f(t)|\mathrm{d}t=\Theta/2$. A physical example here is 
$f(t)\!=\!-g\mu_{\mathrm{B}}B\cos(\omega t)/2$~\cite{Madsen2021} with $g$ the Lande factor, $\mu_{\mathrm{B}}$ the 
electron magnetic moment and $B\cos(\omega t)$ a periodic magnetic field. Due to the fact $|\int_0^t f(t_1)\mathrm{d}t_1|\leq 
g\mu_{\mathrm{B}}B/(2\omega)$, $\mathcal{S}$ is determined by the ratio between $B$ and $\omega$. The OQSL reads 
$\tau=\arcsin(\frac{\omega\Theta}{g\mu_{\mathrm{B}} B})/\omega$, and the optimal states are the states in the $xy$ plane. 
It is obvious that $\tau\leq\pi/(2\omega)$ as $\arcsin(\cdot)$ is always less than or equal to $\pi/2$. This upper bound is nothing 
but the time when the first degenerate point occurs, which leads to an interesting phenomenon that \emph{all targets can be fulfilled 
before the first degenerate point occurs} with the states in the $xy$ plane. In the case that a bounded control $u(t)$ ($|u(t)|\!\leq\! u_b$) 
is invoked, $f(t)$ becomes $u(t)\!-\!g\mu_{\mathrm{B}}B\cos(\omega t)/2$ and the upper bound of $|\int^{t}_0 f(t_1)\mathrm{d}t_1|$
can always overcome $\pi/2$ at a long enough time. Hence, in this case $\mathcal{S}=\{\vec{r}\,|\alpha\!\in\![\Theta/2,
\pi-\Theta/2]\}$ and the OQSL satisfies $\int^{\tau}_0 |g\mu_{\mathrm{B}}B\cos(\omega t)/2-u(t)|\mathrm{d}t=\Theta/2$. The
minimum $\tau$ with respect to $u(t)$ (denoted by $\tau_{\min}$) satisfies the equation $g\mu_{\mathrm{B}} B\sin(\omega\tau_{\min})
/(2\omega)+u_b\tau_{\min}=\Theta/2$, and $\tau_{\min}\approx\Theta/(g\mu_{\mathrm{B}} B+2u_b)$ for a small $\omega$. The
calculation details are in the Supplementary Information.

Another practical scenario to apply Theorem 1 is the one-dimensional Ising model with a longitudinal field, where two
boundary conditions (periodic and open) exist. Let us first consider the case of periodic boundary condition, in which
the Hamiltonian reads $H/J=-\sum_{j=1}^n\sigma_j^z \sigma_{j+1}^z-\sum_{j=1}^n g(t)\sigma_j^z$ with $\sigma^z_{n+1}=\sigma^z_1$.
Here $J>0$ is the interaction strength of the nearest-neighbor coupling, and $g(t)$ is a global time-dependent longitudinal
field. $\sigma^z_j$ is the Pauli Z matrix for $j$th spin. The spin number $n\!\geq\! 3$. In this case, the minimum energy is
$-n[1\!+\!|g(t)|]$, and the maximum energy is $n-\eta[2\!-\!|g(t)|]$ when $|g(t)|\!<\!2$ and $n[|g(t)|\!-\!1]$ when
$|g(t)|\!\geq\! 2$. Here $\eta\!:=\![1+(-1)^{n+1}]/2$. If $|g(t)|\!\geq\! 2$ for all time $t$, the OQSL satisfies the
equation $\int^{\tau}_0|g(t)|\mathrm{d}t=\Theta/(2n)$. Due to the fact that $\int^{\tau}_0|g(t)|\mathrm{d}t\geq \int^{\tau}_0
2\mathrm{d}t=2\tau$, one can immediately finds that $\tau\leq\Theta/(4n)$. If $|g(t)|\!<\! 2$ all the time, Eq.~(\ref{eq:OQSL})
reduces to $2\left(n-\eta\right)\tau+\left(n+\eta\right)\int^{\tau}_0|g(t)|\mathrm{d}t=\Theta$. In this case
$\tau\in\big[\frac{\Theta}{4n},\frac{\Theta}{2n-2\eta}\big]$ since $\int^{\tau}_0|g(t)|\mathrm{d}t\in[0,2\tau]$. For a
$g(t)$ that is not always bounded by $2$, the integration in Eq.~(\ref{eq:OQSL}) needs to be calculated part by part and the
rigorous solution may not easy to be acquired in general. However, in some cases a good approximation can still be obtained
since $\tau$ is usually small. Take $g(t)=B\cos(\omega t)$ as an example, where $B$ and $\omega$ are the amplitude and
frequency. In this case, if $\omega$ is not very large, then $\tau\approx\Theta/[2(n-\eta)+B(n+\eta)]$ when $B\!<\!2$ and
$\tau\approx\Theta/(2Bn)$ when $B\!\geq\! 2$, which are nothing but the OQSLs with respect to the constant field $g(t)=B$.

%========================================= Figure =========================================
\begin{figure*}[tp]
\centering\includegraphics[width=14cm]{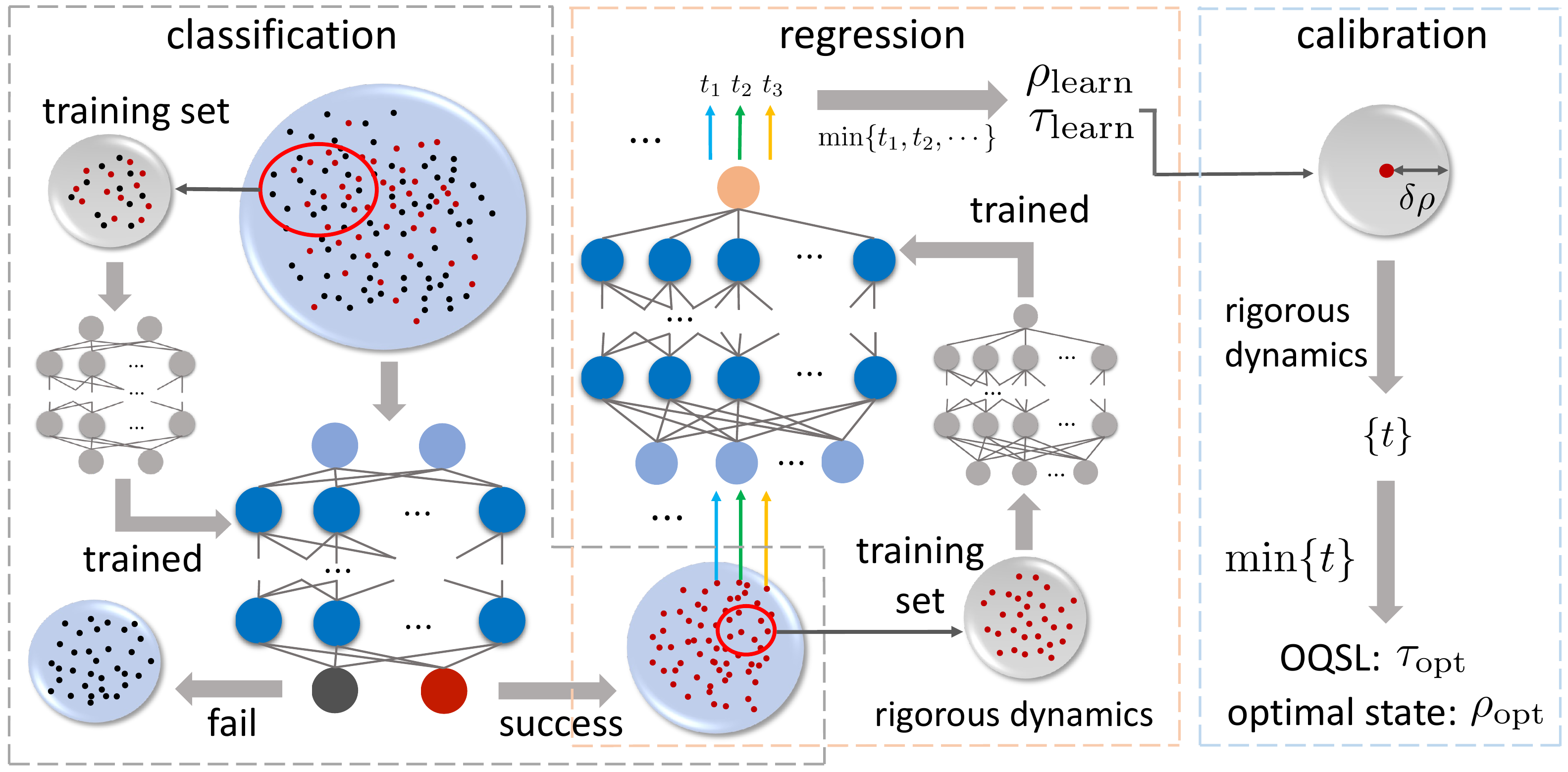}
\caption{CRC methodology to learn the OQSL for complex dynamics. The three steps are
classification (gray box), regression (orange box), and calibration (blue box).}
\label{fig:methodology}
\end{figure*}
%===========================================================================================

In the case of open boundary condition, the Hamiltonian reads $-\sum_{j=1}^{n-1}\sigma_j^z\sigma_{j+1}^z-\sum_{j=1}^n 
g(t)\sigma_j^z$. The minimum energy is $-n[1\!+\!|g(t)|]\!+\!1$, and the maximum energy is $n\!+\!\eta|g(t)|\!-\!1$ when 
$|g(t)|\!\leq\! 1$, $n\!-\!(2\!-\!\eta)[2\!-\!|g(t)|]\!+\!1$ when $|g(t)|\!\in\!(1,2)$, and $n[|g(t)|-1]+1$ when 
$|g(t)|\!\geq\! 2$. For $g(t)\!=\!B\cos(\omega t)$ with a not very large $\omega$, an interesting phenomenon occurs when 
$B\!<\! 2$ and $n$ is even. The OQSL in this case approximates to $\Theta/[n(B\!+\!2)\!-\!2]$ when $B\!\leq\! 1$, and 
$\Theta/[n(B+2)+2(B-2)]$ when $B\!\in\! (1,2)$, which are different from the OQSL under the periodic boundary condition. 
These two OQSLs, as well as their difference, are quite robust to global and local dephasing. Therefore, the OQSL may be 
used to detect whether an even-numbered spin ring is ruptured, especially when the number is not very large. More details 
are in the Supplementary Information.

\vspace{2ex}
\textbf{CRC methodology}

The brute-force search is the most common method for the numerical evaluation of OQSL and is easy to execute for simple 
dynamics. However, when the evaluation of dynamics for one state is too time-consuming, the entire brute-force search would 
be impossible to finish as it usually requires executing thousand and even million rounds of dynamics. In recent years, 
machine learning has been successfully applied to quantum physics for the simulation of complex dynamics, such as the 
theoretical dynamics of many-body systems~\cite{Carleo2017,Hartmann2019,Schmitt2020} and realistic dynamics of experimental 
systems~\cite{Flurin2020,Sivak2022}. With the help of trained neural networks, the computing time to evaluate the dynamics 
significantly reduces compared to the rigorous calculation. Therefore, such learning techniques could be powerful tools to 
evaluate the OQSL. Hereby we provide a three-step methodology (CRC methodology) based on learning to evaluate the OQSL 
for complex dynamics. The three steps are (1) classification; (2) regression; and (3) calibration, as illustrated in 
Fig.~\ref{fig:methodology}. As a matter of fact, classification and regression are two terminologies in supervised learning. 
Classification is a problem to identify the categories of objects and regression is to predict some values related to the 
objects.

The reachable state set $\mathcal{S}$ is crucial in the evaluation of OQSL. It is not only essential for the further
calculation of OQSL, but also reveals information that whether a state is capable to fulfill the target. Hence,
the first step (classification) in CRC methodology is to find $\mathcal{S}$. In this step, a reasonable number of
quantum states and corresponding binary labels ($0$ or $1$) consist of the training set. Quantum states and binary
labels are the input and output of the neural network. In our calculation, label $1$ ($0$) represents the state is
in (not in) $\mathcal{S}$. The performance of the trained network can be tested via a test set. After the training
and performance verification, a large number of random states are input into the network to construct $\mathcal{S}$
according to the outputs. In the following the learned reachable state set in this step is denoted by
$\mathcal{S}_{\mathrm{learn}}$.

The second step is regression. In this step, a subset of $\mathcal{S}_{\mathrm{learn}}$ and the corresponding time to
reach the target consist of the training set. The time to reach the target is extracted from the rigorous dynamics.
Notice that it is possible some states in this subset cannot fulfill the target and need to be removed from the training
set since $\mathcal{S}_{\mathrm{learn}}$ could be slightly different from $\mathcal{S}$ in practice. After the training
and performance verification, all states in $\mathcal{S}_{\mathrm{learn}}$ will be input into the trained network, and
the minimum output ($\tau_{\mathrm{learn}}$) and corresponding states ($\rho_{\mathrm{learn}}$) are extracted. 
The performance of $\tau_{\mathrm{learn}}$ relies on the performance of the trained neural network in this process. 
Usually enlarging the scale of the training set is a possible way to improve the performance of learning. However, in many 
cases this improvement is not always positively correlated to the scale growth of the training set. In the meantime, choosing 
an appropriate neural network would also be helpful, yet whether a network is appropriate usually needs to be thoroughly 
tested case by case. Moreover, large-scale models or quantum machine learning are also possible candidates to further 
improve the performance of $\tau_{\mathrm{learn}}$, and we will continue to investigate this problem in the future. 

In principle $\tau_{\mathrm{learn}}$ could be treated as an approximation of OQSL. However, if the methodology stops here
then the accuracy of learned OQSL would be strongly affected by the residuals, namely, the differences between the true and
predicted values. In the meantime, $\rho_{\mathrm{learn}}$ may not be the actual optimal state in the neighborhood due to
the existence of residuals. To further improve the methodology's performance, we introduce the third step: calibration. In
this step, a reasonable region around $\rho_{\mathrm{learn}}$ in the state space is picked, and the dynamics of enough random
states in this region are calculated rigorously. Then the minimum time to reach the target in this region ($\tau_{\mathrm{opt}}$)
and corresponding state ($\rho_{\mathrm{opt}}$) are picked out. $\tau_{\mathrm{opt}}$ is the final evaluated value of OQSL in
the methodology. Due to the fact that the process of calibration is designed to reduce the influence of residuals, a general principle 
for a proper region in calibration is that in this region it should clearly show that whether $\rho_{\mathrm{learn}}$ is a local 
minimum point. 

To verify the validity of CRC methodology, we apply it in the Landau-Zener model where the reachable state set and OQSL have
been thoroughly discussed via brute-force search among about one million states~\cite{Shao2020}, and thus the methodology's
performance is easy to be tested. The Hamiltonian for the Landau-Zener model is $H=\Delta\sigma_{x}+vt\sigma_z$ with $\Delta$
and $v$ two time-independent parameters. In the step of classification, three training sets with different numbers of data are
used to train the network and about one million states are used as the test set. The scores (correctness of prediction) are
no less than $99.59\%$, $97.83\%$, and $98.00\%$ for all training sets in the cases of $\Delta=0$, $1$, and $2$. In the step
of regression, the mean square errors of learning are on the scale of $10^{-5}$ for $\Delta=0$, $2$, and no larger than
$1.22\times 10^{-4}$ for $\Delta=1$. In the last step, the region for calibration is chosen as $[\alpha_{\mathrm{learn}}\!-\!0.1,
\alpha_{\mathrm{learn}}\!+\!0.1]$ and $[\phi_{\mathrm{learn}}\!-\!0.1,\phi_{\mathrm{learn}}\!+\!0.1]$ where
$\alpha_{\mathrm{learn}}$ and $\phi_{\mathrm{learn}}$ are the spherical coordinates of $\rho_{\mathrm{learn}}$,
i.e., $\cos(\alpha_{\mathrm{learn}})=\mathrm{Tr}(\rho_{\mathrm{learn}}\sigma_z)$ and $\cos(\phi_{\mathrm{learn}})=
\mathrm{Tr}(\rho_{\mathrm{learn}}\sigma_x)/\sin(\alpha_{\mathrm{learn}})$. The results of calibration show that in this case
$\rho_{\mathrm{learn}}$ is just $\rho_{\mathrm{opt}}$ for all values of $\Delta$, and the corresponding $\tau_{\mathrm{opt}}$
coincides with the exact OQSL obtained from the brute-force search. The validity of CRC methodology is then verified. 

%======================================== Figure ========================================
\begin{figure}[tp]
\centering\includegraphics[width=8.5cm]{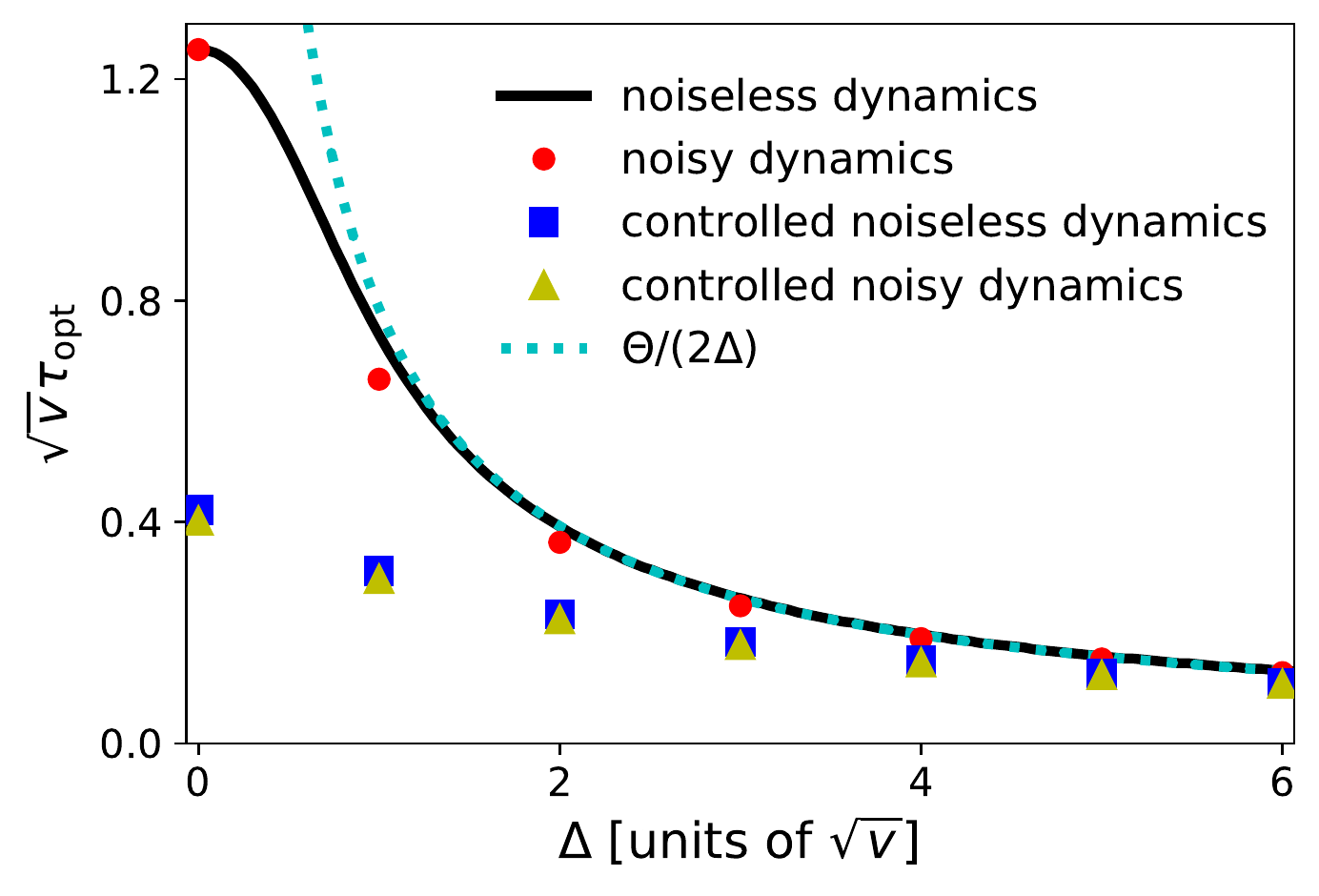}
\caption{OQSL as a function of $\Delta$ in the cases of noiseless dynamics (solid black
line), noisy dynamics (red circles), controlled noiseless dynamics (blue squares), and
controlled noisy dynamics (yellow triangles). The cyan dotted line represents $\Theta/(2\Delta)$.
The target $\Theta=\pi/2$.}
\label{fig:LZdelta}
\end{figure}
%========================================================================================

One advantage of CRC methodology is that it can deal with controlled dynamics, where the brute-force-search
evaluation is usually difficult to realize due to the complexity of twofold optimizations. In the meantime, CRC
methodology can also deal with noisy scenarios where the rigorous dynamics is usually more time-consuming
than the unitary counterpart. Let us still consider the Landau-Zener model with the time-varying control Hamiltonian
$\vec{u}(t)\cdot\vec{\sigma}$. Here $\vec{u}=(u_x(t),u_y(t),u_z(t))$ is the vector of control amplitudes and 
$\vec{\sigma}=(\sigma_x,\sigma_y,\sigma_z)$ is the vector of Pauli matrices. All control amplitudes are assumed
to be in the regime $[-\sqrt{v},\sqrt{v}]$. Both the noiseless and noisy scenarios are studied. In the noisy
scenario, the dynamics is governed by the master equation $\partial_t\rho=-i[H,\rho]+\gamma(\sigma_z\rho\sigma_z-\rho)$
with $\gamma$ the decay rate, which is taken as $0.5\sqrt{v}$ as a demonstration. In this example, the evaluation
of OQSL for $\Delta=0$ via brute-force search among one million states on a daily-use computer costs more than $830$
days, which reduces to $30$ days when the CRC methodology is applied~\cite{footnote}. The result of CRC methodology
shows that all states in the state space can fulfill the target $\Theta=\pi/2$ under control in both noisy and
noiseless cases. Furthermore, the OQSL is very robust to the dephasing in both noncontrolled and controlled cases,
as shown in Fig.~\ref{fig:LZdelta}. In the meantime, the controls can significantly reduce the OQSL when $\Delta$ is
not very large. However, this improvement becomes limited with the increase of $\Delta$. An interesting phenomenon is
that regardless of the existence of both noise and controls, the OQSL always converges to $\Theta/(2\Delta)$, which
is nothing but the OQSL for the Hamiltonian $\Delta\sigma_x$ in the absence of noise~\cite{Shao2020}. This phenomenon
on speed limit is difficult to be revealed by lower-bound-type QSLs not only due to their dependence on both initial
states and time, but also the lousy attainability when controls are involved.

%======================================== Figure ========================================
\begin{figure}[tp]
\centering\includegraphics[width=8.5cm]{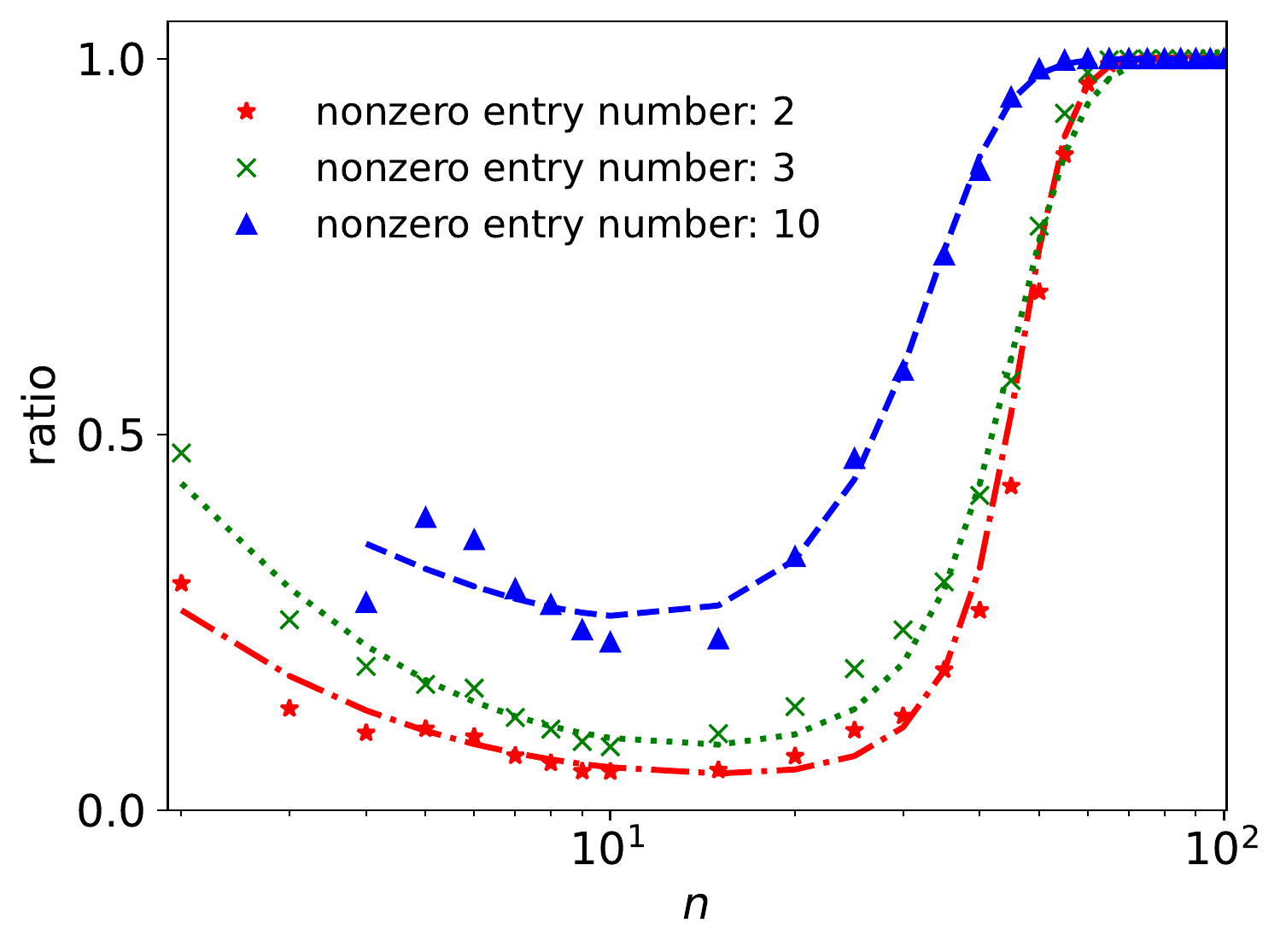}
\caption{Ratio of states that can fulfill the target $\Theta=\pi/2$ in different categories.
The red pentagrams, green crosses, and blue triangles represent the ratios for the states
with $2$, $3$, and $10$ nonzero entries. The dash-dotted red, dotted green, and dashed blue
lines represent the corresponding fitting functions.}
\label{fig:ratio}
\end{figure}
%========================================================================================

Another example we studied is the transverse Ising model with a periodic external field. The Hamiltonian is
$H/J\!=\!-\!\sum_{j=1}^n\sigma_j^z\sigma_{j+1}^z\!-\!\sum_{j=1}^n g(t)\sigma_j^x$ with $g(t)\!=\!B\cos(\omega t)$. In the
demonstration, the amplitude $B$ is taken as $0.5$ and the frequency $\omega/J=1$. Because of the enormous state space
($2^n$), it is difficult to construct a training set that is general enough for the CRC methodology, especially when $n$
is large. To feasibly apply the CRC methodology, we need to analyze the state structure first and reduce the state space
for the study. A simple way to categorize the states is based on the number of nonzero entries in a certain basis, such
as the basis $\{\ket{\uparrow},\ket{\downarrow}\}^{\otimes n}$ considered as follows. $\ket{\uparrow}$ ($\ket{\downarrow}$)
is the eigenstate of $\sigma_z$ with respect to the eigenvalue $1$ ($-1$). Moreover, here we only consider the noiseless
dynamics and that $\mathcal{Q}$ is the set of pure states. The ratios of reachable states for the target $\Theta=\pi/2$ in the 
categories of $2$ (red pentagrams), $3$ (green crosses), and $10$ nonzero entries (blue triangles) are given in Fig.~\ref{fig:ratio}. 
The ratio in each category is obtained from $2000$ random states. It can be seen that basically all states in each category can 
fulfill the target when $n$ is large, which is reasonable as more target directions exist when the dimension is high. Moreover, the 
ratio increases with the rise of the nonzero entry number. More interestingly, the ratio in each category basically fits the function 
$1/(1+an^b e^{-cn^d})$, and the parameters $a,b,c,d$ can be found in the Supplementary Information. The general behaviors of 
the ratio and the physical mechanism behind it are still open questions that require further investigation. The minimum time to 
reach the target for all states in each category is also investigated and the specific results are given in the Supplementary Information, 
which indicates that in this example we only need to focus on the states with few nonzero entries for the study of OQSL.

Next we perform the CRC methodology in the case of $n=10$. The methodology is applied to the categories of states with
$2$ to $5$ nonzero entries. Here we present the result in the category of 2 nonzero entries, and others are given in
the Supplementary Information. $22500$ and $7500$ states and corresponding labels are used as the training and test
sets for the classification. The best score of the trained network we obtained is $94.55\%$. Then about one
million states are input into this network, and the result shows that $7.71\%$ states can fulfill the target, close to
the result ($5.15\%$) obtained from $2000$ random states. In the regression process, $22500$ and $7500$ states consist 
of the training and test sets. The best mean square error is $8.95\!\times\!10^{-4}$ and the corresponding 
$\tau_{\mathrm{learn}}$ is $0.24$, close to the true evolution time ($0.19$) of $\rho_{\mathrm{learn}}$. About $10000$ 
states in the neighborhood of $\rho_{\mathrm{learn}}$ are used in the calibration and the final result is $0.18$. Combing 
the results of the other three categories, the final value of OQSL obtained from the CRC methodology is $0.18$, which can 
be realized by certain states with $2$ nonzero entries.

\vspace{2ex}
\textbf{\large{Methods}}

In both cases of controlled Landau-Zener model and transverse Ising model, 22500 and 7500 datasets are generated for training 
and testing in the classification and regression processes. Each dataset is composed of the initial state and corresponding 
time to reach the given target. In these datasets, the initial states are generated randomly and the time is solved via 
rigorous dynamics. In the case of controlled Landau-Zener model, the optimal control is obtained via the automatic 
differentiation. In the case of transverse Ising model, the initial states are expressed by the matrix product state which 
is implemented via Julia package ITensors~\cite{Fishman2022}, and the time for reaching the given target is calculated with 
time evolving block decimation technique. In the process of calibration, $10000$ datasets are generated in a reasonable 
neighborhood of $\rho_{\mathrm{learn}}$. 

The Python package sklearn~\cite{Pedregosa2011} is used in this paper to build and train the neural networks for the 
classification and regression processes. In the cases of noncontrolled and controlled Landau-Zener models, the layer number 
of the neural network is $5$ to $6$, and each layer contains about 250 neurons. The hyperbolic tangent function and rectified 
linear unit function are chosen as the activation loss function in the classification and regression, respectively. In the 
case of transverse Ising model, the neural networks in classification for the states with $2$, $3$, $4$, and $5$ nonzero 
entries are all activated by the hyperbolic tangent function. With respect to the regression, the activation loss function 
for the neural networks is rectified linear unit function for the states with $2$ nonzero entries, logistic function for 
those with $3$ nonzero entries, and identity function for those with $4$ and $5$ nonzero entries. 

In the process of classification, average cross-entropy loss function is used to train the neural networks, which is of the 
form
\begin{align}
f(\hat{x},x,W)=& -\frac{1}{m}\sum^{m}_{i=0}\left[x_i\ln \hat{x}_i+(1-x_i)\ln(1-\hat{x}_i)\right] \nonumber \\
& +\frac{\alpha}{2m}||W||^2_2,
\end{align}
where $x$ and $\hat{x}$ represent the true results and the results predicted by the neural network. $m$ is the number of 
datasets. $W$ is the weight matrix of the neural network and $\alpha ||W||^2_2=\alpha \sum_{ij}W^2_{ij}$ represents the 
penalty term. And in the regression, the loss function in the training is the mean square error function, 
\begin{equation}
f(t_{\mathrm{pre}},t_{\mathrm{ext}},W)=\frac{1}{2m}\sum^{m}_{i=0} 
\left[t^{(i)}_{\mathrm{pre}}-t^{(i)}_{\mathrm{ext}}\right]^2+\frac{\alpha}{2m}||W||^2_2,
\end{equation}
here $t_{\mathrm{pre}}$ and $t_{\mathrm{ext}}$ are the time predicted by the regression neural network and the exact time 
obtained via rigorous dynamics. More details of the methods can be found in the Supplementary Information. 

When the value of $\Theta$ is changed, the reachable state set changes accordingly, which means all the neural networks in the 
CRC methodology have to be retrained.  How to train general neural networks that work for all target values is still a very challenging  
problem, and requires further and continuous investigations in the future.

\vspace{2ex}
\textbf{\large{Data availability}}

The data that support the findings of this study are available from J.L. upon reasonable request.

\vspace{2ex}
\textbf{\large{Code availability}}

The code used in this study is available from J.L. upon reasonable request.

\vspace{2ex}
\textbf{Acknowledgments}

The authors thank Yuqian Xu for helpful discussion. This work was supported by the National Natural Science Foundation
of China (Grant No.\,12175075).

\vspace{2ex}
\textbf{Author contributions}

J.L. conceived the idea and wrote the manuscript. M.Z. and H.M.Y. performed the calculations. All authors contributed 
to the discussion and reviewed the manuscript.

\vspace{2ex}
\textbf{Competing interests}

The authors declare no competing interests.

\appendix

\section{Connections between different tools to define the target}

It is well known that there exist various types of tools in the quantum speed limit (QSL) to define the target, and different tools may 
lead to different mathematical bounds or methods for the description of QSL. However, it is possible that a physical target could be 
quantified via different tools, and thus these tools should present connections on values. For example, assume the target is defined by 
the tool $\Phi_1$ and a specific value (denoted by $\phi$) of it is taken as the target, then this target can also be represented by 
another tool $\Phi_2$ as long as $\Phi_1$ and $\Phi_2$ have certain connections on values. Denote this connection as a function, i.e., 
$\Phi_2=f(\Phi_1)$, then with the tool of $\Phi_2$ the target value can be expressed by $f(\Phi_1=\phi)$. When the connection is a 
one-to-one correspondence, namely, $f$ is an univalent function, the targets $\phi$ and $f(\Phi_1=\phi)$ are equivalent. In the case 
that $f$ is a multivalent function, this conversion would lead to several different target values. 

\begin{figure}[tp]
\centering\includegraphics[width=8.cm]{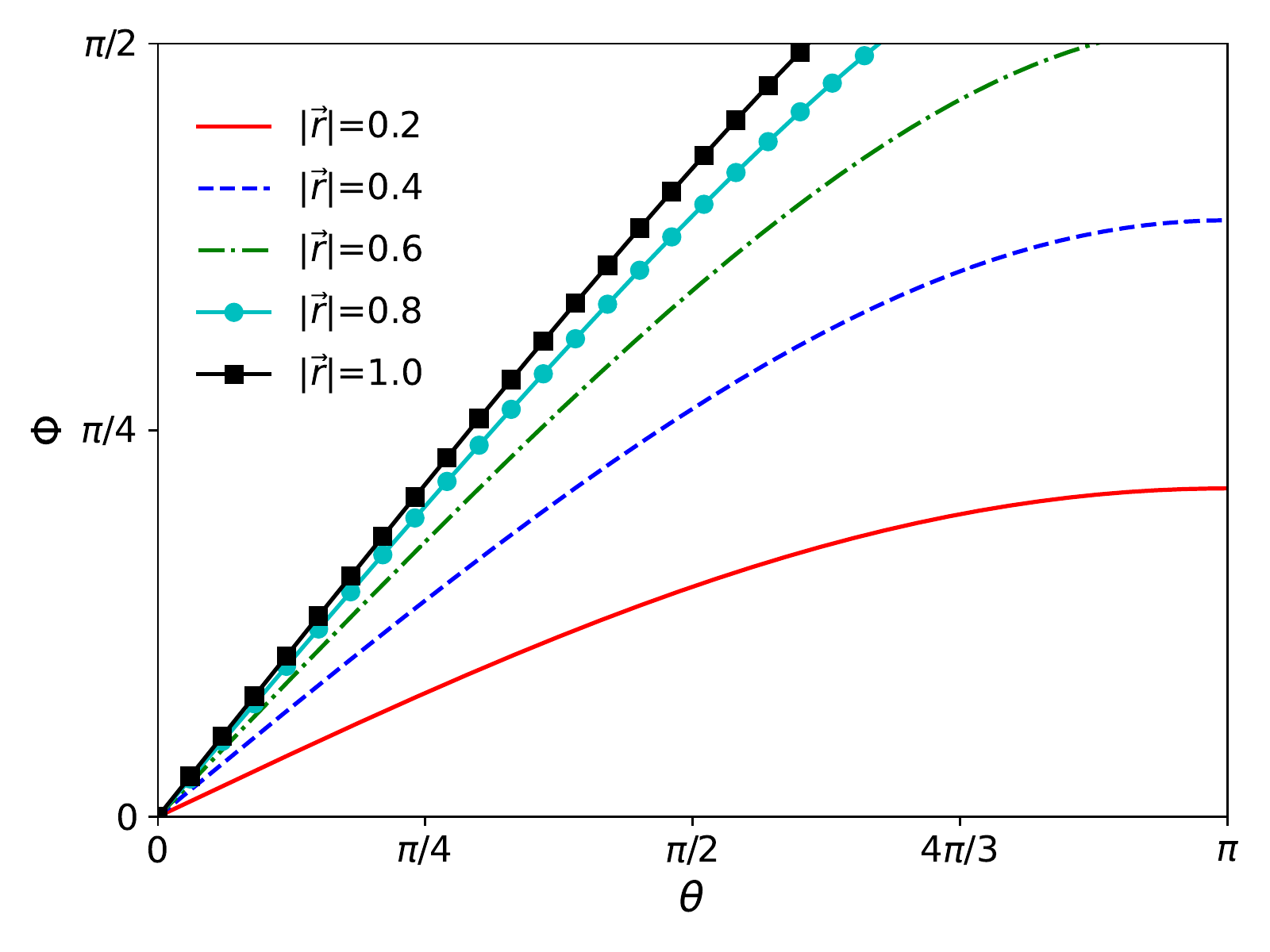}
\caption{Demonstration of the one-to-one correspondence between the angle 
of relative purity and Bloch angle for the states with $|\vec{r}|=0.2$ (solid-red 
line), $|\vec{r}|=0.4$ (dashed-blue line), $|\vec{r}|=0.6$ (dash-dotted-green line), 
$|\vec{r}|=0.8$ (solid-cyan-circle line), and $|\vec{r}|=1.0$ (solid-black-square 
line), respectively. $N=4$ in the plot.}
\label{fig:toolequiva}
\end{figure}

Now we discuss the relation between the angle of relative purity and Bloch angle as a demonstration. In the perspective of QSL, the 
angle ($\Phi\in (0,\pi/2]$) of relative purity could be defined by
\begin{equation}
\Phi = \arccos\left(\frac{\mathrm{Tr}(\rho \rho(t))}{\mathrm{Tr}(\rho^2)}\right),
\label{eq:apx_repurity}
\end{equation}
where $\rho$ is the initial state and $\rho(t)$ is the corresponding evolved state at time $t$. In the Bloch representation, $\rho$ can be 
expressed by
\begin{equation}
\rho = \frac{1}{N}\left(\openone+\sqrt{\frac{N(N-1)}{2}}\vec{r}\cdot\vec{\lambda}\right),
\label{eq:apx_dmBloch}
\end{equation}
where $N$ is the dimension of $\rho$, $\vec{\lambda}$ is the vector of SU($N$) generators and its $i$th entry $\lambda_i$ and $j$th 
entry $\lambda_j$ satisfies $\mathrm{Tr} (\lambda_i \lambda_j)=2\delta_{ij}$ with $\delta_{ij}$ the Kronecker delta function. $\vec{r}$ 
is the Bloch vector satisfying $|\vec{r}|\leq1$. $\openone$ is the identity matrix. Substituting Eq.~(\ref{eq:apx_dmBloch}) into 
Eq.~(\ref{eq:apx_repurity}), one can obtain 
\begin{equation}
\frac{\mathrm{Tr}(\rho\rho(t))}{\mathrm{Tr}(\rho^2)}=\frac{1+(N-1)\vec{r}\cdot\vec{r}(t)}{1+(N-1)|\vec{r}|^2}.
\label{eq:apx_tooltemp}
\end{equation}
In the perspective of QSL, the Bloch angle $\theta\in(0,\pi]$ is defined as the angle between the vectors $\vec{r}$ and $\vec{r}(t)$. Hence,  
Eq.~(\ref{eq:apx_tooltemp}) can be rewritten into 
\begin{equation}
\cos\Phi=\frac{1+(N-1)|\vec{r}|^2\cos\theta}{1+(N-1)|\vec{r}|^2}, 
\label{eq:apx_toolrelation}
\end{equation}
where $\cos\theta\geq \max \big\{\!\!-1,-\frac{1}{(N-1)|\vec{r}|^{2}}\big\}$. This condition is to guarantee that the right-hand term in the 
equation is non-negative.  As demonstrated in Fig.~\ref{fig:toolequiva}, for the same initial state $\theta$ and $\Phi$ has a one-to-one 
correspondence relation and thus they are equivalent on values. 

%====================== Figure ==============================
%===========================================================

In the perspective of the operational definition of quantum speed limit (OQSL), this one-to-one correspondence means that the set of 
target states for the same initial state are exactly the same for these two tools, and hence the reachable state sets of them are also the 
same, indicating that they are actually the same problem. For the tools that no one-to-one correspondence exists, such as the Bures angle 
and Bloch angle, the reachable state sets are not exactly the same for these tools and the result of OQSL may not be the same. 

\section{Connection between the OQSL and the quantum brachistochrones problem}

The OQSL has a deep connection with the quantum brachistochrones problem. In the problem of quantum brachistochrones, people 
usually concern about the minimum time for a given initial state to a target state $\rho_{\mathrm{tar}}$ or the realization of a certain gate.  
Due to the fact that in OQSL the initial state has been optimized, instead of a given initial state, with the OQSL we can study the minimum 
time for a set of initial states (denoted by $\mathcal{Q}$) to reach the target state. Notice that if $\rho_{\mathrm{tar}}\in \mathcal{Q}$, 
the optimal state in $\mathcal{Q}$ to reach $\rho_{\mathrm{tar}}$ must be $\rho_{\mathrm{tar}}$ itself for any Hamiltonian and the 
corresponding time is nothing but zero, indicating that this is a trivial case. Therefore,  here we only consider the nontrivial case that 
$\rho_{\mathrm{tar}} \notin \mathcal{Q}$ is satisfied. Next we take a qubit case as a demonstration. 

Consider a noncontrolled Hamiltonian $H=\omega\sigma_z/2$ with $\sigma_z$ the Pauli Z matrix and $\omega$ the energy difference. 
The other two Pauli matrices are denoted by $\sigma_x$ and $\sigma_y$. The target state $\rho_{\mathrm{tar}}=(\ket{0}-\ket{1})/\sqrt{2}$ 
with $\ket{0}$ ($\ket{1}$) the eigenstate of $\sigma_z$ corresponding to the eigenvalue $1$ ($-1$).  The set $\mathcal{Q}=\{\rho|
\mathrm{Tr}(\rho\sigma_x)\geq 0\}$.  In this case, the reachable state set can be written as 
\begin{equation}
\mathcal{S}=\{\rho|\rho\in \mathcal{Q}~\&~e^{-iHt}\rho e^{iHt}=\rho_{\mathrm{tar}}, \exists t\}.
\end{equation}

In the Bloch representation with $\ket{0}$ the north pole [$\vec{r}=(r_x,r_y,r_z)^{\mathrm{T}}$], $\mathcal{Q}$ can be rewritten into 
$\mathcal{Q}=\{\vec{r}|r_x\geq 0\}$ and the equation $e^{-iHt}\rho e^{iHt}=\rho_{\mathrm{tar}}$ can be rewritten into 
\begin{equation}
\begin{cases}
\frac{1}{2}(1+r_z)=\frac{1}{2}, \\
\frac{1}{2}e^{-i\omega t}(r_x-i r_y)=-\frac{1}{2}.
\end{cases}
\label{eq:apx_bb1}
\end{equation}
To make sure that these equations have legitimate solutions of time, $\vec{r}$ has to satisfies the conditions $r_z=0$ and $r^2_x+r^2_y=1$. 
Hence, $\mathcal{S}$ can be expressed by 
\begin{equation}
\mathcal{S}=\{\vec{r}\,|\,r_z=0, r_x\geq 0, r^2_x+r^2_y=1\}.
\end{equation}
Utilizing the spherical coordinates of $\vec{r}$, i.e., 
\begin{equation}
\vec{r}=\eta (\sin\alpha\cos\varphi,\sin\alpha\sin\varphi,\cos\alpha)^{\mathrm{T}} 
\end{equation}
with $\alpha\in[0,\pi]$ and $\varphi\in[0,2\pi)$, $\mathcal{S}$ can be rewritten into 
\begin{equation}
\mathcal{S}=\left\{\vec{r}\,|\,\eta=1,\alpha=\pi/2,\varphi\in[0,\pi/2]\cup[3\pi/2,2\pi)\right\}.
\end{equation}
The solution of time for Eq.~(\ref{eq:apx_bb1}) is 
\begin{equation}
t = \frac{(2k+1)\pi-\varphi}{\omega}, k=0,1,2\cdots.
\end{equation}
The minimum time $\tau=\pi/(2\omega)$, which can be attained by the state $(0,1,0)^{\mathrm{T}}$. 

This result is quite reasonable from the perspective of geometry. As a matter of fact, $\mathcal{S}$ is nothing but half of the $xy$ plane 
with $r_x\geq 0$. Due to the fact that the dynamics is the rotation about $z$ axis, the state that can reach the target state 
$ (-1,0,0)^{\mathrm{T}}$ in the minimum time to is just the $y$ axis, i.e., $(0,1,0)^{\mathrm{T}}$. 

\section{The OQSL for time-dependent Hamiltonian with time-independent eigenstates}

\subsection{Proof of the Theorem}
\label{sec:apx_proof}

Consider the time-dependent Hamiltonian of the form
\begin{equation}
H(t)=\sum_i E_i(t)\ket{E_i}\bra{E_i}, \label{eq:apx_Ht}
\end{equation}
where the energies are assumed to be ordered ascendingly, i.e., $E_0(t)\leq E_1(t)\leq \cdots\leq E_{N-1}(t)$
(not all the equalities are saturated simultaneously) and $\ket{E_i}$ is independent of the time for any
subscript $i$. With this Hamiltonian, the OQSL $\tau$ satisfies
\begin{equation}
\int_0^\tau E_{N-1}(t)-E_{0}(t)\mathrm{d}t=\Theta,
\end{equation}
where $E_{N-1}(t)$ and $E_{0}(t)$ are the highest and lowest energies of the Hamiltonian at time $t$. The proof is as follows.

In the case of unitary dynamics, any SU($N$) generator satisfies $U(t)\lambda_i U^{\dagger}(t)=\sum_j C_{ij}(t) \lambda_j$ 
with $U(t)$ a unitary operator, then the dynamics of $\vec{r}$ can be written as $\vec{r}(t) = C^{\mathrm T}(t) \vec{r}$. 
Due to the polar decomposition, $C^{\mathrm{T}}$ can be decomposed into $C^{\mathrm{T}}=OS$ with $O$ 
a real orthogonal matrix and $S$ a real positive semi-definite symmetric matrix. Hence, $C^{\mathrm{T}}$ represents a 
deformation of the Bloch sphere along principal axes determined by $S$ and then a proper rotation due to $O$~\cite{Nielsen2000}. 
Recall that $\mathrm{Tr} (\lambda_i \lambda_j)=2\delta_{ij}$ with $\delta_{ij}$ the Kronecker delta function, then $C_{ij}(t)$ 
can be further solved as $C_{ij}(t)=\mathrm{Tr} (U(t)\lambda_i U^{\dagger}(t) \lambda_j)/2$.

With the Hamiltonian~(\ref{eq:apx_Ht}), the unitary operator can be expressed by
\begin{eqnarray}
U(t) &=& e^{-i\sum_m\int_0^t E_m(t_1)\mathrm{d}t_1\ket{E_m}\bra{E_m}}\nonumber \\
&=& \sum_m e^{-i\int_0^t E_m(t_1) \mathrm{d}t_1} \ket{E_m}\bra{E_m},
\end{eqnarray}
which indicates
\begin{equation}
C_{ij}(t)=\frac{1}{2} \sum_{mn} e^{i\int_0^t E_m(t_1)-E_n(t_1) \mathrm{d}t_1}
[\lambda_i]_{mn}^{*} [\lambda_j]_{mn}
\end{equation}
with $[\lambda_j]_{mn}$ the $mn$th entry of $\lambda_j$. In the energy basis
$\{\ket{E_0},\ket{E_1}, \dots,\ket{E_{N-1}}\}$, $C(t)$ has the same
structure with the time-dependent Hamiltonian~\cite{Shao2020}, i.e.,
\begin{equation}
C(t) = \bigoplus\limits_{n=1}^{N-1} V(n,t),
\label{eq:apx_Ct}
\end{equation}
where $V(n,t)=\left[\bigoplus\limits_{i=0}^{n-1}M(\Delta_{ni})\right]\bigoplus 1$
with
\begin{equation}
M(x) =\left(\begin{array}{cc}
\cos x & -\sin x \\
\sin x & \cos x  \\
\end{array}\right)
\end{equation}
and $\Delta_{ni}=\int_0^t E_n(t_1)-E_i(t_1) \mathrm{d}t_1$. Then the angle between the
initial and evolved Bloch vectors is
\begin{equation}
\cos\theta=\frac{\vec{r}(t)\cdot\vec{r}}{|\vec{r}|^2}=\frac{\vec{r}^{\mathrm{T}}
C(t)\vec{r}} {|\vec{r}|^2}.
\end{equation}
Utilizing Eq.~(\ref{eq:apx_Ct}), it can be further calculated as
\begin{equation*}
\cos\theta \!=\! 1-\frac{1}{|\vec{r}|^2}\!\sum_{n=1}^{N-1}\!\sum_{i=0}^{n-1}
[1\!-\!\cos(\Delta_{ni})](r_{n^2+2i-1}^2 \!+\! r_{n^2+2i}^2)
\end{equation*}
with $r_i$ the $i$th element of $\vec{r}$. Hence, the set $\mathcal S$ can be
directly expressed by
\begin{eqnarray}
\mathcal{S} &=& \Big\{\vec{r}\,\big|\,1-\cos\Theta=\frac{1}{|\vec{r}|^2}
\sum_{n=1}^{N-1}\sum_{i=0}^{n-1}[1-\cos(\Delta_{ni})] \nonumber \\
& & \times \left(r_{n^2+2i-1}^2+r_{n^2+2i}^2\right), \exists t \Big\}.
\end{eqnarray}

To further obtain the OQSL, the two-step proof strategy used in Appendix B in Ref.~\cite{Shao2020}
needs to be applied. Define
\begin{equation*}
f(t):=\frac{1}{|\vec{r}|^2}\sum_{n=1}^{N-1}\sum_{i=0}^{n-1}
[1-\cos (\Delta_{ni})](r_{n^2+2i-1}^2+r_{n^2+2i}^2).
\end{equation*}
Substituting the equation
\begin{equation}
\int_0^{\tau} E_{N-1}(t)-E_{0}(t)\mathrm{d}t=\Theta
\end{equation}
into the expression of $f(t)$, it can be seen that
\begin{equation}
\frac{\partial f(t)}{\partial t}\Big|_{t=\tau}\geq 0,
\end{equation}
which indicates $\tau$ is in the first monotonic increasing regime of $f(t)$. In the meantime,
it can also be found that $f(\tau)\leq 1-\cos\Theta$, which is due to the fact  
\begin{align}
f(\tau)& \leq(1-\cos\Theta)\sum_{n=1}^{N-1}\sum_{i=0}^{n-1}
\frac{r_{n^2+2i-1}^2+r_{n^2+2i}^2}{|\vec{r}|^2} \nonumber \\
& \leq 1-\cos\Theta.
\end{align} 
Here the inequality $1-\cos(\Delta_{ni}(\tau))\leq 1-\cos\Theta$ has been applied. If the solution of 
$f(t)=1-\cos\Theta$ is not in the first increasing regime of $f(t)$, then $t$ is obviously larger than 
$\tau$; if this solution is in the first increasing regime, then due to $f(\tau)\leq 1-\cos\Theta=f(t)$ 
one can also see that $t\geq \tau$. Hence, $\tau$ is a lower bound of the time to reach the target angle.

Now we discuss the optimal probe states to reach the OQSL. To let the equation $1-\cos\Theta=f(\tau)$
holds, the term $r^2_{n^2+2i-1}+r^2_{n^2+2i}$ for the subscripts $n$, $i$ satisfying $\Delta_{ni}\neq
\int^{\tau}_0 E_{N-1}(t)-E_0(t)\mathrm{d}t$ has to vanish. Further assume the degeneracy of the ground
states and highest excited states are $p$ and $q$, namely, $E_0(t)=E_1(t)=\dots=E_{p-1}(t)$ and
$E_{N-q}(t)=E_{N-q+1}(t)=\dots=E_{N-1}(t)$, then it is easy to see that $r^2_{n^2+2i-1}+r^2_{n^2+2i}$
can only be nonzero when $n\in [N-q,N-1]$ and $i\in [0,p-1]$, which indicates that the optimal state
is of the form
\begin{equation}
\sum^{N-1}_{i=0}\frac{1}{N}\ket{E_i}\bra{E_i}+\!\!\!\sum_{k\in [0,p-1],\atop l\in [N-q,N-1]}
\!\!\!\xi_{kl}\ket{E_k}\bra{E_l} +\xi_{kl}^{*} \ket{E_l}\bra{E_k},
\label{eq:apx_optstate}
\end{equation}
where $\xi_{kl}=\sqrt{\frac{N-1}{2N}}(r_{l^2+2k-1}-ir_{l^2+2k})$.
In the energy basis $\{\ket{E_0},\ket{E_1},\dots,\ket{E_{N-1}}\}$, the state above can be written as
\begin{equation}
\frac{1}{N}\openone+\left(\begin{array}{ccc}
0 & 0 & \xi \\
0 & \cdots & 0 \\
\xi^{\dagger} & 0 & 0
\end{array}\right),
\label{eq:apx_optstate1}
\end{equation}
where $\openone$ is a $N$-dimensional identity matrix, and $\xi$ is a $p$ by $q$ matrix with $kl$th entry
$\xi_{kl}$. To make sure the density matrix is positive-semidefinite, according to the Schur complement theorem
$\xi$ needs to satisfy
\begin{equation}
\xi^{\dagger}\xi\leq \frac{1}{N^2}\openone_{q},
\end{equation}
where $\openone_{q}$ is a $q$-dimensional identity matrix. The theorem is then proved. \hfill $\blacksquare$

\subsection{Example: two-level systems} \label{sec:apx_twolevel}

Here we take a two-level system as a demonstration of the Theorem. Consider the Hamiltonian
\begin{equation}
H(t) = f(t)\sigma_z,
\label{eq:apx_ft_sz}
\end{equation}
where $f(t)$ is a function of time $t$, and $\sigma_z$ is the Pauli Z matrix.
In the Bloch representation, the evolved Bloch vector can be solved as
\begin{eqnarray*}
r_x(t) &=& r_x \cos\left[2\int_0^t f(t_1)\mathrm{d}t_1\right]-
           r_y \sin\left[2\int_0^t f(t_1)\mathrm{d}t_1\right], \\
r_y(t) &=& r_x \sin\left[2\int_0^t f(t_1)\mathrm{d}t_1\right]+
           r_y \cos\left[2\int_0^t f(t_1)\mathrm{d}t_1\right], \\
r_z(t) &=& r_z,
\end{eqnarray*}
where $(r_x, r_y, r_z)^{\mathrm{T}}=\vec{r}$ is the Bloch vector of the initial state. Based on this
dynamics, the angle between the initial and evolved states is
\begin{equation}
\cos\theta=\frac{\cos\left[2\int_0^t f(t_1)\mathrm{d}t_1\right] (r_x^2+r_y^2)+r_z^2}{|\vec{r}|^2},
\end{equation}
which indicates that the time to reach the target angle $\Theta$ satisfies the following equation
\begin{equation}
\sin^2\left[\int_0^t f(t_1)\mathrm{d}t_1\right] = \frac{|\vec{r}|^2}{|\vec{r}|^2-r_z^2}
\sin^2\left(\frac{\Theta}{2}\right).
\label{eq:apx_target}
\end{equation}
Rewrite $\vec{r}$ into
\begin{equation}
\vec{r}=\eta (\sin\alpha\cos\varphi,\sin\alpha\sin\varphi,\cos\alpha)^{\mathrm{T}}
\end{equation}
with $\eta\in [0,1]$, $\alpha\in[0,\pi]$ and $\varphi\in[0,2\pi]$, and Eq.~(\ref{eq:apx_target})
reduces to
\begin{equation}
\sin^2\alpha=\frac{\sin^2\left(\frac{\Theta}{2}\right)}{\sin^2\left[\int_0^t f(t_1)\mathrm{d}t_1\right]}.
\label{eq:apx_target1}
\end{equation}
Now consider that $|\int_0^t f(t_1)\mathrm{d}t_1|$ is upper bounded by $c_{f}$, then in the case that
$c_f<\Theta/2$, $\sin^2\left[\int_0^t f(t_1)\mathrm{d}t_1\right]$ is always less than $\sin^2(\Theta/2)$,
which gives $\sin^2\alpha > 1$. This means no state can fulfill the target as $\sin^2\alpha$ is always
equal or less than 1. In the case that $c_f\in[\Theta/2,\pi/2]$,
\begin{equation}
\sin^2\left[\int_0^t f(t_1)\mathrm{d}t_1\right]\leq \sin^2 c_f\leq 1.
\end{equation}
Hence,
\begin{equation}
\sin^2\alpha\geq\frac{\sin^2\left(\frac{\Theta}{2}\right)}{\sin^2 c_f},
\end{equation}
indicating that the states that can fulfill the target satisfies $\alpha\in[\alpha_{f},\pi-\alpha_{f}]$ with
\begin{equation}
\alpha_f=\arcsin\left(\frac{\sin(\frac{\Theta}{2})}{\sin c_f}\right).
\end{equation}
In the case that $c_f>\pi/2$, $\sin^2[\int_0^t f(t_1)\mathrm{d}t_1]$ can reach all the values between 0 and 1,
and $\sin^2\alpha\geq\sin^2(\Theta/2)$, therefore, the states satisfies $\alpha\in[\Theta/2,\pi-\Theta/2]$.
In a word, the set $\mathcal{S}$ can be expressed by
\begin{equation}
\mathcal{S}=
\begin{cases}
\emptyset, & c_f<\frac{\Theta}{2}, \\
\{\vec{r}\,|\,\alpha\in[\alpha_f,\pi-\alpha_f]\}, & c_f\in[\frac{\Theta}{2},\frac{\pi}{2}], \\
\{\vec{r}\,|\,\alpha\in[\frac{\Theta}{2},\pi-\frac{\Theta}{2}]\}, & c_f >\frac{\pi}{2}.
\end{cases}
\label{eq:apx_S}
\end{equation}
Here $\emptyset$ is the empty set, and in the second and third circumstances $\eta\in (0,1]$ and
$\varphi\in[0,2\pi]$.

With respect to the OQSL, due to the fact that the eigenvalues are always $f(t)$ and $-f(t)$, the
maximum and minimum ones are always $|f(t)|$ and $-|f(t)|$, respectively. Based on the Theorem, the OQSL
$\tau$ then satisfies
\begin{equation}
\int^{\tau}_0 |f(t)|\mathrm{d}t=\frac{\Theta}{2}. \label{eq:apx_ft_tau}
\end{equation}

A physical example for the Hamiltonian~(\ref{eq:apx_ft_sz}) is the energy splitting coming from
Zeeman effect, i.e.,
\begin{equation}
f(t)=-\frac{g\mu_{\mathrm{B}}}{2}B(t)
\end{equation}
with $g$ the Lande factor and $\mu_{\mathrm{B}}$ the electron magnetic moment. $B(t)$ is the
time-dependent magnetic field. For a periodic magnetic field $B(t)=B\cos(\omega t)$ with $B,\omega>0$,
it is easy to see
\begin{equation}
\left|\int_0^t f(t_1)\mathrm{d}t_1\right|=\left|\frac{g\mu_{\mathrm{B}}B}{2\omega}\sin(\omega t)\right|
\leq \frac{g\mu_{\mathrm{B}}B}{2\omega}.
\end{equation}
According to Eq.~(\ref{eq:apx_S}), the set $\mathcal{S}$ reads
\begin{equation}
\mathcal{S}=
\begin{cases}
\emptyset, & \frac{g\mu_{\mathrm{B}}B}{2\omega}<\frac{\Theta}{2}, \\
\{\vec{r}\,|\,\alpha\in[\alpha_f,\pi-\alpha_f]\},
& \frac{g\mu_{\mathrm{B}}B}{2\omega}\in[\frac{\Theta}{2},\frac{\pi}{2}], \\
\{\vec{r}\,|\,\alpha\in[\frac{\Theta}{2},\pi-\frac{\Theta}{2}]\},
& \frac{g\mu_{\mathrm{B}}B}{2\omega} >\frac{\pi}{2}.
\end{cases}
\end{equation}
Here $\eta\in (0,1]$, $\varphi\in[0,2\pi]$ and
\begin{equation}
\alpha_f=\arcsin\left(\frac{\sin(\frac{\Theta}{2})}{\sin(\frac{g\mu_{\mathrm{B}}B}{2\omega})}\right).
\end{equation}

Utilizing the Theorem, Eq.~(\ref{eq:apx_ft_tau}) can be written as
\begin{equation}
\int^{\tau}_0 |\cos(\omega t)|\mathrm{d}t=\frac{\Theta }{g\mu_{\mathrm{B}}B}.
\end{equation}
In the case that $\frac{g\mu_{\mathrm{B}}B}{2\omega}<\frac{\Theta}{2}$, $\tau=\infty$ as
no states can reach the target. Hence we only consider the non-trivial case that
$\frac{g\mu_{\mathrm{B}}B}{2\omega}\geq\frac{\Theta}{2}$, which means
$\frac{\Theta}{g\mu_{\mathrm{B}}B}\leq\frac{1}{\omega}$, and therefore
$\int^{\tau}_0 |\cos(\omega t)|\mathrm{d}t\leq 1/\omega$, namely, $\int^{\tau}_0 |\cos(\omega t)
|\mathrm{d}(\omega t)\leq1$. The integration of $|\cos(\omega t)|$ is only less or equal to 1
when $\omega t\leq\pi/2$, in which regime $\cos(\omega t)$ is always non-negative, hence, the
integration is equivalent to be performed on $\cos(\omega t)$. Finally, the equation above can be
rewritten into
\begin{equation}
\int^{\tau}_0 \cos(\omega t)\mathrm{d}t=\frac{\Theta }{g\mu_{\mathrm{B}}B},
\end{equation}
which immediately gives the analytical expression of $\tau$ as below
\begin{equation}
\tau = \frac{1}{\omega} \arcsin \left(\frac{\omega\Theta}{g\mu_{\mathrm{B}} B}\right).
\end{equation}
An interesting fact in this case is that the first degenerate point shows at $t=\pi/(2\omega)$,
and the OQSL is always less or equal to this time, indicating that the target $\Theta$, regardless
of its value, can always be reached before this first degeneracy point.

Next we consider a controlled case that
\begin{equation}
f(t)=-\frac{g\mu_{\mathrm{B}}}{2}B\cos(\omega t)+u(t),
\end{equation}
where $|u(t)|\leq u_b$ is a bounded control. Since
\begin{align}
& \left|\int^t_0 -\frac{g\mu_{\mathrm{B}}}{2}B\cos(\omega t_1)+u(t_1)\mathrm{d}t_1 \right| \nonumber \\
= & \left|\frac{g\mu_{\mathrm{B}}B}{2\omega}\sin(\omega t)-\int^t_0 u(t_1)\mathrm{d}t_1\right| \nonumber \\
\leq & \frac{g\mu_{\mathrm{B}}B}{2\omega}+\left|\int^t_0 u(t_1)\mathrm{d}t_1\right| \nonumber \\
\leq & \frac{g\mu_{\mathrm{B}}B}{2\omega}+u_b t,
\end{align}
which can be larger than $\pi/2$ for a long enough time, in this case
\begin{equation}
\mathcal{S}=\left\{\vec{r}~\big|~\alpha\in\left[\frac{\Theta}{2},\pi-\frac{\Theta}{2}\right]\right\}.
\end{equation}
The OQSL here satisfies
\begin{equation}
\int^{\tau}_0 \left|\frac{g\mu_{\mathrm{B}}B}{2}\cos(\omega t)-u(t)\right|\mathrm{d}t=\frac{\Theta}{2}.
\end{equation}
Then the minimum value of $\tau$ (denoted by $\tau_{\min}$) can be solved via the problem
\begin{eqnarray*}
\tau_{\min} &=& \min_{u(t)} \tau, \nonumber \\
& & \mathrm{subject~to}~\begin{cases}
\int^{\tau}_0 |\frac{g\mu_{\mathrm{B}}B}{2}\cos(\omega t)\!-\!u(t) |\mathrm{d}t=\frac{\Theta}{2}, \\
|u(t)| \leq u_b.
\end{cases}
\end{eqnarray*}
This problem can be solved by maximizing the function $|\frac{1}{2}g\mu_{\mathrm{B}}B\cos(\omega t)-u(t)|$
under the constraint that its integration is fixed. Since $\cos(\omega t)$ is a monotonic function within
the regime $[0,\pi/(2\omega)]$, one could have
\begin{align}
& \int^{\frac{\pi}{2\omega}}_0 \left|\frac{g\mu_{\mathrm{B}}B}{2}\cos(\omega t)-u(t)\right|\mathrm{d}t \nonumber \\
\leq & \int^{\frac{\pi}{2\omega}}_0 \left[\frac{g\mu_{\mathrm{B}}B}{2}\cos(\omega t)+u_b\right]\mathrm{d}t \nonumber \\
= & \frac{g\mu_{\mathrm{B}}B}{2\omega}+\frac{\pi}{2\omega}u_b.
\end{align}
Notice the condition to make sure $\mathcal{S}\neq \emptyset$ is $\frac{g\mu_{\mathrm{B}}B}{2\omega}\geq
\frac{\Theta}{2}$. In this case, the upper bound of $\int^{\frac{\pi}{2\omega}}_0 |\frac{g\mu_{\mathrm{B}}B}{2}
\cos(\omega t)-u(t)|\mathrm{d}t$ is larger than $\Theta/2$, indicating that the integration will reach
$\Theta/2$ before the time $\pi/(2\omega)$ with proper controls. Hence, $\tau_{\min}$ must be less than
$\pi/(2\omega)$. Under this condition, the maximum value of $|\frac{1}{2}g\mu_{\mathrm{B}}B\cos(\omega t)-u(t)|$
is attained when $u(t)\equiv -u_b$ due to the fact that $\cos(\omega t)$ is a monotonic function here. Therefore,
$\tau_{\min}$ satisfies the equation
\begin{equation}
\frac{g\mu_{\mathrm{B}} B}{2\omega}\sin(\omega\tau_{\min})+u_b\tau_{\min}=\frac{\Theta}{2}.
\end{equation}
If $\omega$ is small, $\tau_{\min}$ approximates to
\begin{equation}
\tau_{\min}\approx \frac{\Theta}{g\mu_{\mathrm{B}} B+2u_b}.
\end{equation}

\subsection{Example: one-dimensional Ising model with a longitudinal field}
\label{sec:longtitude_Ising}

\subsubsection{Periodic boundary condition}

In the following we consider the one-dimensional Ising model with a longitudinal field. The Hamiltonian of this system reads
\begin{equation}
H/J=-\sum_{j=1}^n\sigma_j^z \sigma_{j+1}^z-\sum_{j=1}^n g(t)\sigma_j^z,
\label{eq:apx_longitudinal}
\end{equation}
where $J>0$ is the interaction strength of the nearest-neighbor coupling, and $g(t)$ is a global time-dependent longitudinal
field. $\sigma^z_j$ is the Pauli Z matrix for $j$th spin. The Hamiltonian satisfies the periodic boundary condition
$\sigma^z_{n+1}=\sigma^z_1$. Here we only consider the case that $n\geq 3$.

Now we calculate the maximum and minimum eigenvalues of $H/J$. Since the Hamiltonian only
contains the Pauli Z matrix, it is naturally a diagonal matrix in the space consisting of
the eigenspaces of $\sigma^z_j$ for all $j$. Denote $\ket{\uparrow_j}$ and $\ket{\downarrow_j}$
as the eigenstates of $\sigma^z_j$ with respect to the eigenvalues $1$ and $-1$, then the eigenvalues
of $-\sigma^z_j\sigma^z_{j+1}-g(t)\sigma^z_j$ are $1+g(t)$, $1-g(t)$, $-1-g(t)$, and $-1+g(t)$,
and the corresponding eigenstates are $\ket{\downarrow_j \uparrow_{j+1}}$,
$\ket{\uparrow_j \downarrow_{j+1}}$, $\ket{\uparrow_j \uparrow_{j+1}}$,
and $\ket{\downarrow_j \downarrow_{j+1}}$. The eigenvalues of $H/J$ can be obtained by the
summation of a certain number of these four terms. For instance, the eigenvalue with respect to the
eigenstate $\ket{\uparrow}^{\otimes n}$ is $\sum^n_{j=1}[-1-g(t)]=n[-1-g(t)]$.

Regarding the minimum eigenvalue of $H/J$, it is easy to see that the minimum eigenvalue of
$-\sigma^z_j\sigma^z_{j+1}-g(t)\sigma^z_j$ is $-1-g(t)$ when $g(t)\geq 0$ and $-1+g(t)$ when $g(t)\leq 0$,
namely, $-1-|g(t)|$. Therefore, the minimum eigenvalue of $H/J$ is
\begin{equation}
E_{\min,\mathrm{p}}=-n\left[1+|g(t)|\right],
\end{equation}
which can be attained by the eigenstate $\ket{\uparrow}^{\otimes n}$ when $g(t)\geq 0$ and
$\ket{\downarrow}^{\otimes n}$ when $g(t)\leq 0$.

%====================== Figure ==============================
\begin{figure}[tp]
\centering\includegraphics[width=8.cm]{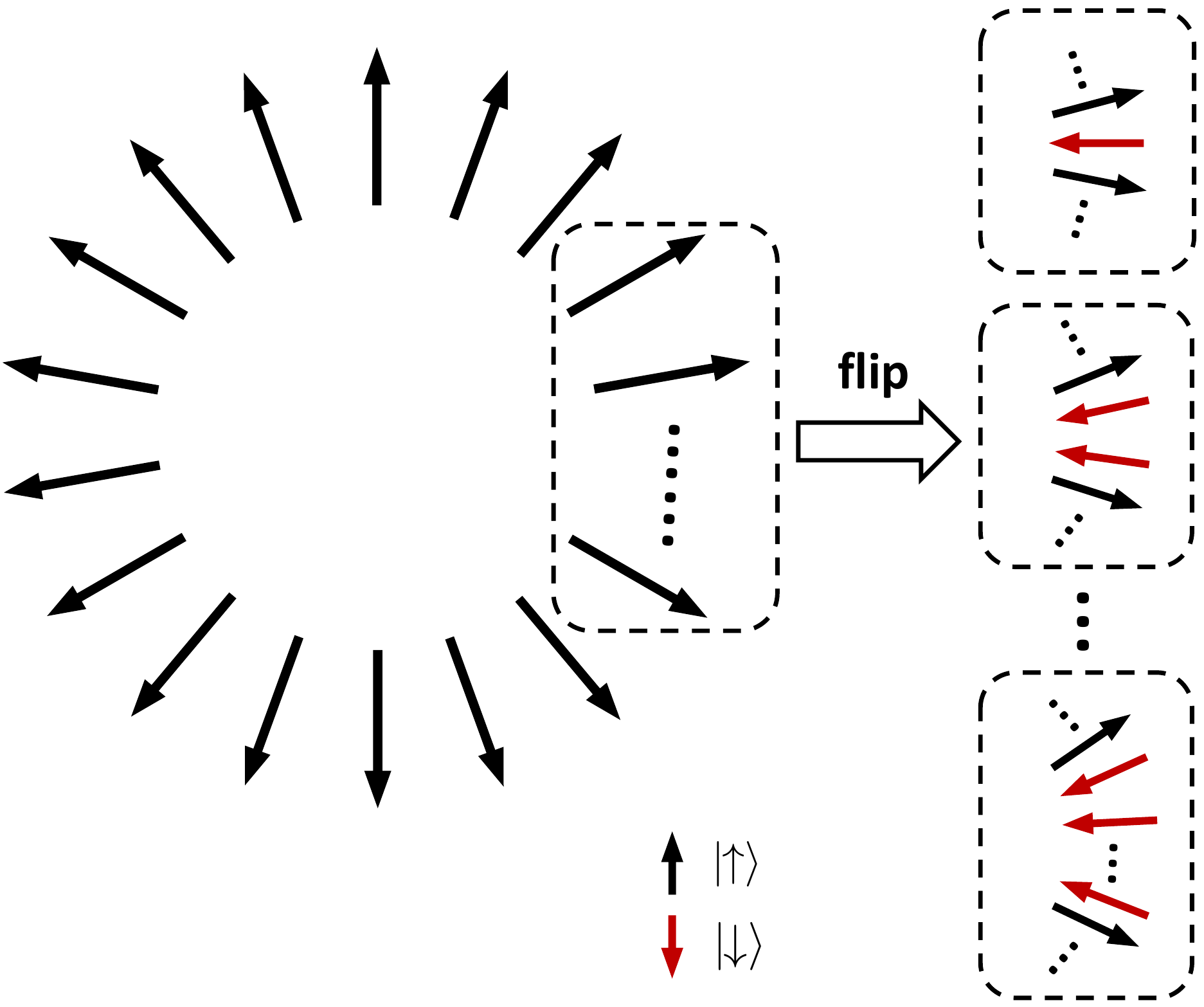}
\caption{Schematic of obtaining any eigenstate of $H/J$ by
flipping any number of $\ket{\uparrow}$ (black up arrow)
into $\ket{\downarrow}$ (red down arrow) in the state
$\ket{\uparrow}^{\otimes n}$.
\label{fig:longitudeIsing}}
\end{figure}
%===========================================================

Next, we calculate the maximum eigenvalue. For an eigenstate $\otimes^{n}_{j=1}\ket{a_j}$
($a_j\!\!=\,\uparrow,\downarrow$), denote the number of $\ket{\uparrow_{j}\uparrow_{j+1}}$,
$\ket{\downarrow_{j}\downarrow_{j+1}}$, $\ket{\downarrow_{j}\uparrow_{j+1}}$, and
$\ket{\uparrow_{j}\downarrow_{j+1}}$ ($j\in [1,N]$) are $x_1$, $x_2$, $x_3$, and $x_4$, respectively.
For example, for the state $\ket{\uparrow\downarrow\uparrow}$, $x_1=1$, $x_2=0$, and $x_3=x_4=1$.
Notice that any eigenstate of $H/J$ can be obtained by flipping any number of $\ket{\uparrow}$ in the
state $\ket{\uparrow}^{\otimes n}$ into $\ket{\downarrow}$. As long as the number of flipped
spins is less than $n$, no matter how many spins are flipped, there always exists a pair of
$\ket{\uparrow\downarrow}$ and $\ket{\downarrow\uparrow}$ at the boundary of the flipped spins,
as shown in Fig.~\ref{fig:longitudeIsing}. For example, assume a flip occurs at the $j$th spin and $k$ spins
are flipped. Then the state of the $(j-1)$th and $j$th spins must be $\ket{\uparrow_{j-1}\downarrow_{j}}$,
and that of the $(j+k-1)$th and $(j+k)$th spins must be $\ket{\downarrow_{j+k-1}\uparrow_{j+k}}$. If all
the spins are flipped, no $\ket{\uparrow\downarrow}$ and $\ket{\downarrow\uparrow}$ exist in the state.
The simultaneous existence of $\ket{\uparrow\downarrow}$ and $\ket{\downarrow\uparrow}$ in the flip
indicates that $x_3$ always equals to $x_4$. Utilizing $x_1$, $x_2$, $x_3$, and the condition $x_1+x_2+2x_3=n$,
the eigenvalue of $H/J$ can be expressed by $-[2+g(t)]x_1+[-2+g(t)]x_2+n$. Hence, the calculation of the maximum
eigenvalue is equivalent to a linear optimization problem: the maximization of $-[2+g(t)]x_1+[-2+g(t)]x_2+n$ under
some constraints on $x_1$, $x_2$, and $x_3$. It is easy to see that the natural constraints on $x_1$, $x_2$, and $x_3$
are $0\leq x_1,x_2\leq n$ and $0\leq x_3\leq \lfloor{n/2}\rfloor$. Here
$\lfloor\cdot\rfloor$ is the floor function. Combing the equation $x_1+x_2+2x_3=n$, the condition $0\leq x_3\leq
\lfloor{n/2}\rfloor$ is equivalent to $0\leq x_1+x_2 \leq n$ when $n$ is even and $1\leq x_1+x_2 \leq n$ when
$n$ is odd, which can be unified as $\frac{1}{2}[1+(-1)^{n+1}]\leq x_1+x_2 \leq n$. This condition is fully
contained by the constraint $0\leq x_1, x_2\leq n$. Hence, the full linear optimization problem can be expressed by
\begin{align}
& \max_{x_1,x_2}~-\left[2+g(t)\right]x_1+\left[-2+g(t)\right]x_2+n,  \nonumber \\
& \mathrm{subject}~\mathrm{to}~
\begin{cases}
\eta\leq x_1+x_2 \leq n, \\
x_1,x_2\in \mathbb{N}.
\end{cases}
\end{align}
Here $\eta:=\frac{1}{2}[1+(-1)^{n+1}]$ and $\mathbb{N}$ is the set of natural numbers.

To solve this problem, four cases have to be discussed: (1) $g(t)\leq -2$, (2) $-2<g(t)\leq 0$, (3) $0<g(t)<2$,
and (4) $g(t)\geq 2$. In the case that $g(t)\leq -2$, the coefficients $-[2+g(t)]\geq 0$ and $-2+g(t)\leq 0$,
indicating that the maximum eigenvalue is obtained when $x_1$ is largest and $x_2$ vanishes, i.e., $x_1=n$, $x_2=0$.
The corresponding maximum eigenvalue is $n[-g(t)-1]$. In the case that $g(t)\in (-2,0]$, both coefficients
$-[2+g(t)]$ and $-2+g(t)$ are negative, and the maximum eigenvalue is attained by the lower bounds of $x_1$ and $x_2$.
If $n$ is even, the minimum value of $x_1$ and $x_2$ are both zero, which leads to the maximum value $n$. If $n$
is odd, the maximum value is attained by $x_1=1$, $x_2=0$ due to the fact that $-[2+g(t)]$ is larger than $-2+g(t)$.
The corresponding maximum value is $n-2-g(t)$. In the case that $g(t)\in (0,2)$, the situation is similar to
the second one. The maximum value is $n$ and attained by $x_1=x_2=0$ when $n$ is even. For an odd $n$, the maximum
value is $n-2+g(t)$, which can be attained by $x_1=0$, $x_2=1$. In the last case that $g(t)\geq2$, $-[2+g(t)]\leq 0$
and $-2+g(t)\geq 0$. The maximum value is $n[g(t)-1]$, which is attained by $x_1=0$, $x_2=n$. In summary, the
maximum eigenvalue of $H/J$ is of the form
\begin{equation}
E_{\max,\mathrm{p}} = \begin{cases}
n-\eta\left[2-|g(t)|\right], & |g(t)|<2, \\
n\left[|g(t)|-1\right], & {|g(t)|\geq 2}.
\end{cases}
\end{equation}

%============================================ Figure =======================================
\begin{figure}[tp]
\centering\includegraphics[width=8cm]{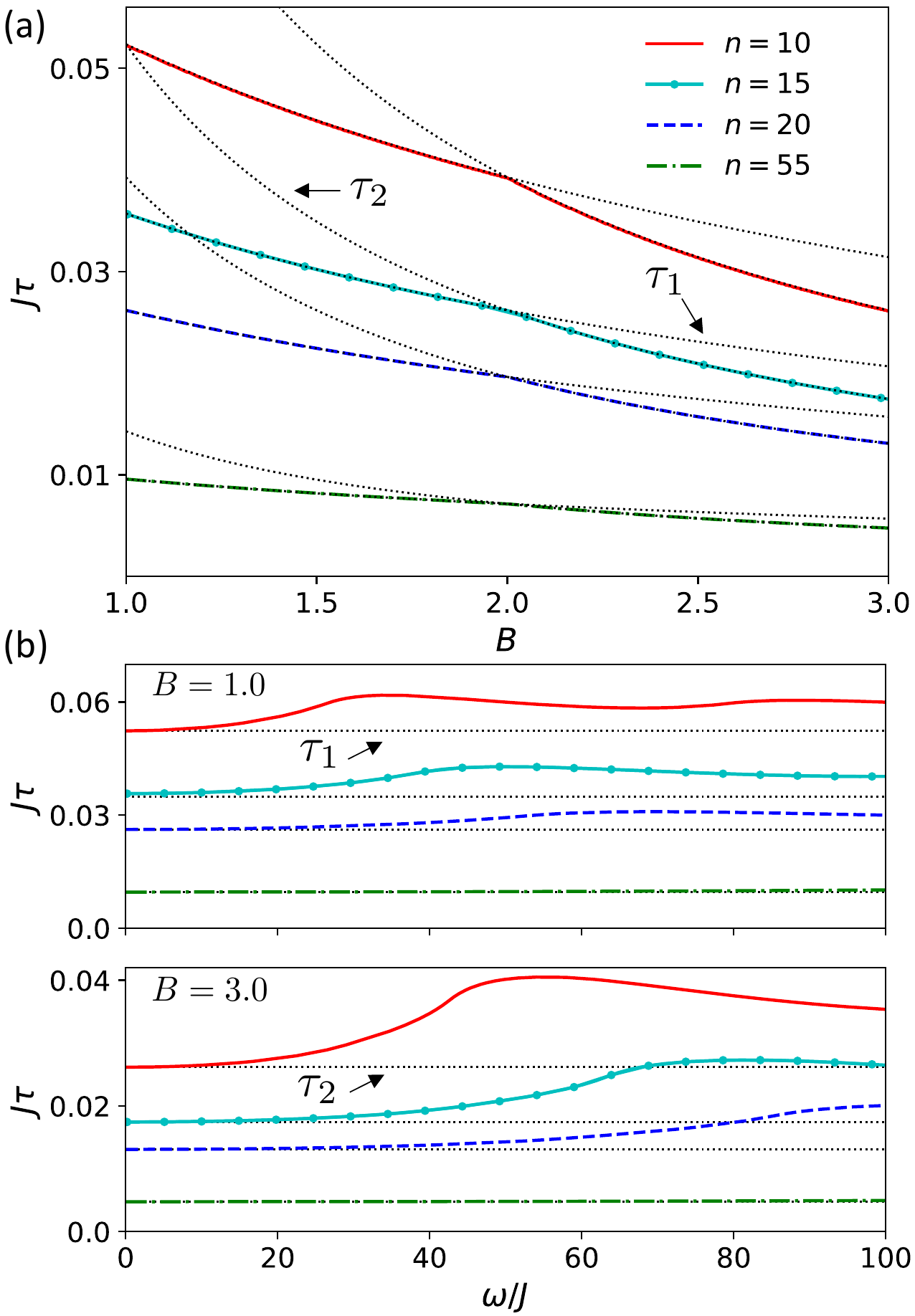}
\caption{Validity of the approximation with the changes of (a) the amplitude $B$ and (b)
the frequency $\omega$ for $n=10$ (solid red lines), $n=15$ (solid circle cyan lines), $n=20$
(dashed blue lines), and $n=55$ (dash-dotted green lines). $\omega/J=1$ in (a), and in (b)
$B=1.0$ and $B=3.0$ for the upper and lower panels, respectively.}
\label{fig:apx_Ising}
\end{figure}
%============================================================================================

Now we consider a specific case that $g(t)=B\cos(\omega t)$, where $B$ is a positive amplitude and $\omega$ is the
frequency. The OQSL $\tau$ is solved via the equation
\begin{equation}
\int^{\tau}_0 E_{\max,\mathrm{p}}(t)-E_{\min,\mathrm{p}}(t)\mathrm{d}t=\Theta.
\label{eq:apx_theorem1}
\end{equation}
When $B<2$, $|g(t)|$ is always less than 2, which means $E_{\max,\mathrm{p}}$ always takes the form
$n-\eta\left[2-|g(t)|\right]$, and Eq.~(\ref{eq:apx_theorem1}) reduces to
\begin{equation}
2\left(n-\eta\right)\tau+\left(n+\eta\right)\int^{\tau}_0|g(t)|\mathrm{d}t=\Theta.
\end{equation}
For a not very large $\omega$, $\int^{\tau}_0|g(t)|\mathrm{d}t=\frac{B}{\omega}\sin(\omega \tau)\approx B\tau$.
Hence,
\begin{equation}
\tau\approx\frac{\Theta}{2(n-\eta)+B(n+\eta)}=:\tau_1.
\label{eq:apx_tau1}
\end{equation}
When $B>2$, the relation between $|g(t)|$ and 2 is not fixed at different time. However, for a not very large $\omega$,
$\tau$ is still very small in this case, which means $E_{\max,\mathrm{p}}$ takes the form $n[g(t)-1]$ before the time
$\tau$, and $\int^{\tau}_0|g(t)|\mathrm{d}t$ still approximates to $B\tau$. Therefore, according to Eq.~(\ref{eq:apx_theorem1}),
$\tau$ approximates to
\begin{equation}
\tau\approx\frac{\Theta}{2Bn}=:\tau_2.
\label{eq:apx_tau2}
\end{equation}

The validity of approximation is numerically tested with the changes of amplitude $B$ and frequency $\omega$ for
different spin number $n$. As shown in Fig.~\ref{fig:apx_Ising}(a), the performance of approximation is very well
for different values of $B$ when $\omega$ is not extremely large [$\omega/J=1$ in the plot]. As to the frequency $\omega$,
the approximation is valid when $\omega$ is no larger than around $10$ for both $B=1.0$ [upper panel in Fig.~\ref{fig:apx_Ising}(b)]
and $B=3.0$ (lower panel). As a matter of fact, $\tau_1$ and $\tau_2$ are nothing but the OQSLs for the constant external field
$g(t)=B$. Hence, the validity of approximation for a large regime of $\omega$ indicates that the OQSL is way more sensitive to
the amplitude than the frequency as long as the frequency is not extremely large.

\subsubsection{Open boundary condition}

Next we consider the case of the open boundary condition. The corresponding Hamiltonian reads
\begin{equation}
H/J=-\sum_{j=1}^{n-1}\sigma_j^z \sigma_{j+1}^z-\sum_{j=1}^n g(t)\sigma_j^z.
\end{equation}
In this case, the minimum eigenvalue of
$-\sigma^z_j\sigma^z_{j+1}-g(t)\sigma^z_j$ is $-1-g(t)$ $[-1+g(t)]$ for $g(t)\geq 0$ $[g(t)\leq 0]$, which leads
to the minimum eigenvalue of $H/J$
\begin{equation}
E_{\min,\mathrm{o}}=-n\left[1+|g(t)|\right]+1.
\end{equation}
The minimum eigenvalue can be attained by the eigenstate $\ket{\uparrow}^{\otimes n}$
$[\ket{\downarrow}^{\otimes n}]$ for $g(t)\geq 0$ $[g(t) < 0]$.

To calculate the maximum eigenvalue, we rewrite the Hamiltonian into the form
\begin{equation}
H/J=H_{\mathrm{p}}+\sigma_{n}^z\sigma_{1}^z,
\end{equation}
where $H_{\mathrm{p}}$ is the Hamiltonian under the periodic boundary condition. Now let us denote $E_{\max,\mathrm{p}}$
and $\ket{E_{\max,\mathrm{p}}}$ as the maximum eigenvalue and corresponding eigenstate of $H_{\mathrm{p}}$,
which is actually already obtained in the previous discussion. Notice that the eigenstates of $H_{\mathrm{p}}$ are
also eigenstates of $\sigma^z_n\sigma^z_1$, and the corresponding eigenvalues can only be $1$ and $-1$. Hence, if
$\ket{E_{\max,\mathrm{p}}}$ also corresponds to the eigenvalue 1, i.e., $\sigma^z_n\sigma^z_1\ket{E_{\max,\mathrm{p}}}
=\ket{E_{\max,\mathrm{p}}}$, then the maximum energy for the entire Hamiltonian is just $E_{\max,\mathrm{p}}+1$.
As a matter of fact, this is just the case for any $n$ in the regime $|g(t)|\geq 2$, and for odd $n$ in the regime
$|g(t)|< 2$. Hence, the maximum eigenvalue $E_{\max}$ for these cases reads
\begin{equation*}
E_{\max,\mathrm{o}}=\begin{cases}
n\left[|g(t)|-1\right]+1, & |g(t)|\geq 2, \\
n+|g(t)|-1, & |g(t)|< 2~\mathrm{and}~n~\mathrm{is}~\mathrm{odd}.
\end{cases}
\end{equation*}

For an even $n$ in the regime $|g(t)|< 2$, $E_{\max,\mathrm{p}}-1=n-1$ may not be the maximum eigenvalue anymore. Another
possible candidate must be among the eigenvalues of which the corresponding eigenstate $\ket{E_{\mathrm{c}}}$ satisfies
$\sigma^z_n\sigma^z_1\ket{E_{\mathrm{c}}}=\ket{E_{\mathrm{c}}}$. It is obvious that we only need to find the maximum
eigenvalues in this case and compare it with $E_{\max,\mathrm{p}}-1$. This maximization problem can still be formulated as
a linear optimization problem as follows
\begin{align}
& \max_{x_1,x_2}~-\left[2+g(t)\right]x_1+\left[-2+g(t)\right]x_2+n+1,  \nonumber \\
& \mathrm{subject}~\mathrm{to}~
\begin{cases}
2\leq x_1+x_2 \leq n, \\
x_1,x_2\in \mathbb{N}, \\
|g(t)|\leq 2.
\end{cases}
\end{align}
The constraint $x_1+x_2\geq 2$ comes from the fact that $\sigma^z_n\sigma^z_1\ket{E_{\mathrm{c}}}=\ket{E_{\mathrm{c}}}$
is equivalent to require $x_1\geq 1$ or $x_2\geq 1$, and $x_1+x_2+2x_3=n$ requires $x_1+x_2$ has to be an even number when
$n$ is even. Hence, $x_1+x_2$ has to be no smaller than $2$. Since both the coefficients $-[2+g(t)]$ and $-2+g(t)$ are
nonpositive in this case, the maximum value must be attained by $x_1=2,x_2=0$ or $x_1=0,x_2=2$. Therefore, in this case
the maximum eigenvalue is $n+2|g(t)|-3$. Next we need to compare the value between $n-1$ and $n+2|g(t)|-3$. As a matter of
fact, it is easy to see when $n-1$ is larger when $|g(t)|<1$ and $n+2|g(t)|-3$ is larger when $|g(t)|>1$. In summary, the
maximum eigenvalue $E_{\max,\mathrm{o}}$ under the open boundary condition reads
\begin{equation}
\begin{cases}
n-1, & |g(t)|\leq 1~\mathrm{and}~n~\mathrm{is}~\mathrm{even}, \\
n+2|g(t)|-3, & 1< |g(t)|< 2~\mathrm{and}~n~\mathrm{is}~\mathrm{even}, \\
n+|g(t)|-1, & |g(t)|< 2~\mathrm{and}~n~\mathrm{is}~\mathrm{odd}, \\
n\left[|g(t)|-1\right]+1, & |g(t)|\geq 2. \\
\end{cases}
\end{equation}
Utilizing the symbol $\eta=[1+(-1)^{n+1}]/2$, the equation above can be rewritten into
\begin{equation}
E_{\max,\mathrm{o}}\!=\!\begin{cases}
n+\eta|g(t)|-1, & |g(t)|\leq 1, \\
n-(2-\eta)[2-|g(t)|]+1, & 1<|g(t)|<2, \\
n\left[|g(t)|-1\right]+1, & |g(t)|\geq 2. \\
\end{cases}
\end{equation}

Next we calculate the OQSL. In the case that $|g(t)|\leq 1$, $\tau$ satisfies the equation
\begin{equation}
(n+\eta)\int_0^{\tau}|g(t)|\mathrm{d}t+(2n-2)\tau=\Theta.
\end{equation}
It is easy to see that here $\int_0^{\tau}|g(t)|\mathrm{d}t$ is less than $\tau$, indicating that
\begin{equation}
\tau\geq \frac{\Theta}{3n-2+\eta}.
\end{equation}
When $1<|g(t)|<2$, $\tau$ satisfies
\begin{equation}
(n+2-\eta)\int_0^{\tau}|g(t)|\mathrm{d}t+(2n-4+2\eta)\tau=\Theta,
\end{equation}
which gives
\begin{equation}
\frac{\Theta}{4n}<\tau<\frac{\Theta}{3n-2+\eta}
\end{equation}
due to the fact that $\tau<\int_0^{\tau}|g(t)|\mathrm{d}t<2\tau$. When $|g(t)|\geq 2$, the OQSL satisfies
\begin{equation}
2n\int_0^{\tau}|g(t)|\mathrm{d}t=\Theta,
\end{equation}
which means $\tau\leq\Theta/(4n)$.

Let us still consider a specific form of $g(t)$ that $g(t)=B\cos(\omega t)$. Similar to the case with the periodic
boundary condition, the approximated expressions of OQSL can also be analytically obtained utilizing the
approximation $\int^{\tau}_0|g(t)|\mathrm{d}t\approx B\tau$ for a not very large $\omega$. In the regime $B\geq 2$,
the OQSL is the same with $\tau_2$ [Eq.~(\ref{eq:apx_tau2})]. A more interesting phenomenon occurs in the regime
$B<2$, where the OQSL is different from $\tau_1$ [Eq.~(\ref{eq:apx_tau1})] for an even $n$. Specifically, the OQSL is
\begin{equation}
\tau\approx\frac{\Theta}{nB+2n-2}=:\tau_{3}
\end{equation}
when $B\leq 1$, and it is
\begin{equation}
\tau\approx\frac{\Theta}{nB+2n+2B-4}
\end{equation}
when $1<B<2$. The maximum gap between the OQSLs for periodic and open boundary conditions happens at the point
$B=0$, i.e., when no external field exists. In this case, the OQSL can be rigorously solved and the difference is
\begin{equation}
\tau_{\mathrm{3}}-\tau_{\mathrm{1}}=\frac{\Theta}{2n(n-1)}=:\Delta\tau.
\end{equation}
The optimal states to realize $\tau_1$ and $\tau_3$ are in the form of Eq.~(\ref{eq:apx_optstate1}). One thing that
should be noticed is that the dimension of $\xi$ in the case of periodic boundary condition could be different from
that in the case of the open boundary condition due to the different degeneracy of minimum and maximum energies in these
two cases.

\subsubsection{Robustness analysis}

The dependence on the boundary condition indicates that the OQSL may be used to detect whether an even-numbered spin
ring is ruptured. To do that, one needs to prepare the optimal states in Eq.~(\ref{eq:apx_optstate}) and then measure
$\mathrm{Tr}(\rho_0\rho_t)$ and $\mathrm{Tr}(\rho^2_t)$ at time $\tau_3$ and $\tau_1$, which can be realized via
techniques like randomized measurements~\cite{Enk2012,Elben2020}. Here $\rho_0$ and $\rho_t$ are the initial state
and evolved state at time $t$. After the measurement, the Bloch angle can be calculated via the equation
\begin{equation}
\cos(\theta(t))=\frac{\mathrm{Tr}(\rho_0\rho_t)-2^{-n}}{\sqrt{\left[\mathrm{Tr}(\rho^2_0)-2^{-n}\right]
\left[\mathrm{Tr}(\rho^2_t)-2^{-n}\right]}}.
\label{eq:apx_isingcos}
\end{equation}
If the target is fulfilled at time $\tau_3$, then the ring is ruptured, and it is complete if the target is fulfilled
at the time $\tau_1$.

A more interesting fact is that the evolution time for the states in Eq.~(\ref{eq:apx_optstate}) is robust to the
global and local dephasing. The global dephasing is described by the master equation
\begin{equation}
\partial_t\rho_t=-i[H,\rho_t]+\gamma_g\left(J_z \rho_t J_z-\frac{1}{2}\left\{\rho_t,J^2_z\right\}\right)
\label{eq:apx_global}
\end{equation}
with $\gamma_g$ the decay rate and $J_z=\frac{1}{2}\sum^n_{j=1}\sigma^z_j$, and the local dephasing is described by
\begin{equation}
\partial_t\rho_t=-i[H,\rho_t]+\sum^n_{j=1}\gamma_{l,j}\left(\sigma^z_j\rho_t\sigma^z_j-\rho_t\right),
\label{eq:apx_local}
\end{equation}
where $\gamma_{l,j}$ is the decay rate for $j$th spin.

%========================================= Figure =========================================
\begin{figure*}[tp]
\centering\includegraphics[width=18cm]{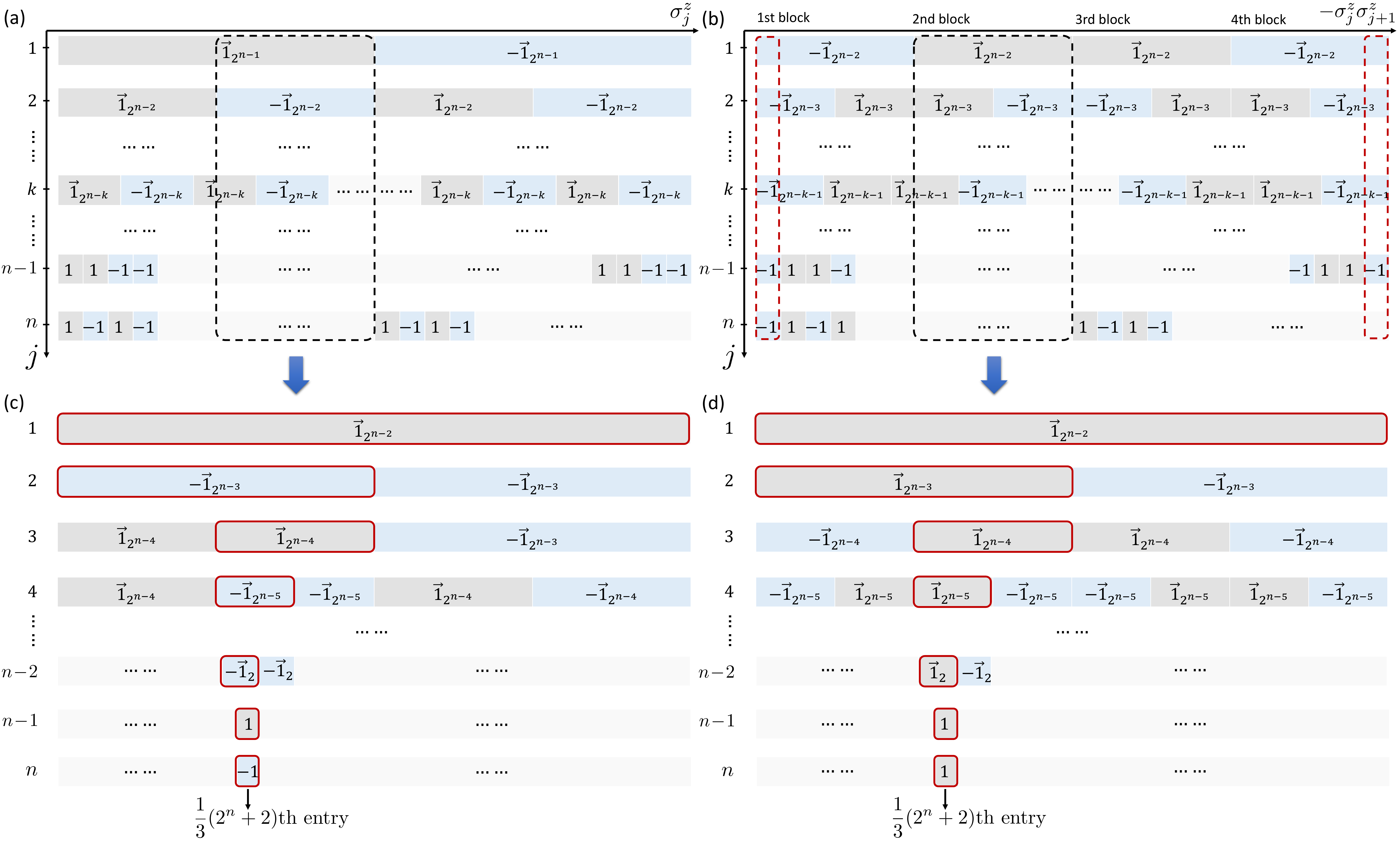}
\caption{Schematic for the search of entry positions of the minimum and maximum energies.
(a) The diagonal entry distribution for $\sigma^z_j$; (b) The diagonal entry distribution
for $-\sigma^z_j\sigma^z_{j+1}$. [(c),(d)] The second blocks for $\sigma^z_j$ and
$-\sigma^z_j\sigma^z_{j+1}$ for the search of the maximum energy.}
\label{fig:entrysearch}
\end{figure*}
%==========================================================================================

Now we analytically discuss this robustness under global and local dephasing. We need to emphasize that the optimal
states [Eq.~(\ref{eq:apx_optstate})] in the noiseless case may not keep optimal when global and local dephasing are
involved, and the corresponding evolution time to reach the target may also not be the OQSL anymore. The analysis of OQSL
under the noise requires the CRC methodology. Here we only discuss the robustness of the evolution time for the states in
Eq.~(\ref{eq:apx_optstate}).

Recall that the states in Eq.~(\ref{eq:apx_optstate}) can be written into Eq.~(\ref{eq:apx_optstate1}) in the basis
$\{\ket{E_0},\ket{E_1},\cdots,\ket{E_{2^n-1}}\}$. Without the external field, the degeneracy of ground states
and the highest energy levels are both
two. In the meantime, due to the fact that $\sigma^z_j$ (for any $j$) and $J_z$ are both diagonal in this basis, we are
allowed to denote $J_z=\mathrm{diag}(A,\dots,G)$ with $A$ and $G$ 2-dimensional diagonal matrices, and $\sigma^z_j=
\mathrm{diag}(C_j,\dots,D_j)$ with $C_j$ and $D_j$ 2-dimensional diagonal matrices. Utilizing these notations, the master
equation for global dephasing [Eq.~(\ref{eq:apx_global})] reduces to the evolution of the block $\xi$ as follows
\begin{equation}
\partial_t \xi_t=i(E_{\max}-E_{\min})\xi_t+\gamma_g A\xi_t G-\frac{\gamma_g}{2}\left(\xi_t G^2+A^2\xi_t\right),
\label{eq:apx_xig}
\end{equation}
where $\xi_t$ is the evolved block at time $t$, and the one for local dephasing [Eq.~(\ref{eq:apx_local})] reduces to
\begin{equation}
\partial_t \xi_t=i(E_{\max}-E_{\min})\xi_t+\sum_j\gamma_{l,j}\left(C_j\xi_t D_j-\xi_t\right).
\label{eq:apx_xil}
\end{equation}
As long as the specific forms of $A$, $G$, $C_j$, and $D_j$ are known, the dynamics can be easily solved. Next, we show
the calculations of these blocks.

It is not difficult to see that $\sigma^z_j$ is easy to be expressed in the basis $\{\ket{\uparrow},
\ket{\downarrow}\}^{\otimes n}$, and the specific forms of $\sigma^z_j$ (diagonal values) for different values of
$j$ are shown in Fig.~\ref{fig:entrysearch}(a), where $\vec{1}_k$ ($-\vec{1}_k$) represents a $k$-dimensional vector with
all entries $1$ ($-1$). To find the expressions of $C_j$ and $D_j$, we need to know the entry positions of minimum and maximum
energies for the Hamiltonian $-\sum_j\sigma^z_j\sigma^z_{j+1}$ and extract the values of $\sigma^z_j$ in the same positions
to reconstruct $C_j$ and $D_j$. The expression of $-\sigma^z_j\sigma^z_{j+1}$ in the basis $\{\ket{\uparrow},
\ket{\downarrow}\}^{\otimes n}$ for different values of $j$ are given in Fig.~\ref{fig:entrysearch}(b). In this
diagram, searching the entry positions of the minimum and maximum energies is equivalent to searching a column with the
most number of $-1$ and $1$. It can be seen that the entries of $-\sigma^z_j\sigma^z_{j+1}$ for all values of $j$ are
symmetric, indicating that the entire diagram can be divided into four blocks, where the first and fourth (second and third)
blocks are mirror symmetric. The positions with respect to the minimum energy are easy to locate since only the first and
last entries of $-\sigma^z_j\sigma^z_{j+1}$ are always $-1$ for all values of $j$. Hence, their summation (summation of the
column in dashed-red boxes) would also be the minimum. In the meantime, the first and last entries of $\sigma^z_j$ are always
$1$ and $-1$ for all values of $j$, indicating that $C_j=\sigma_z$. Moreover, due to the fact that $J_z$ is half of the
summation of all $\sigma^z_j$, the entry positions in $J_z$ that correspond to the minimum energy are also the first and
last entries, which means $A=\mathrm{diag}(n/2,-n/2)=n\sigma_z/2$.

For the sake of finding the entry positions of the maximum energy, we need to locate the position where the entry is
always $1$ for any value of $j$, namely, a column in the diagram where all entries are $1$. It is obvious that it can
only exist in the second and third blocks. Due to the symmetry, we only need to consider the second block. As shown in
Fig.~\ref{fig:entrysearch}(d), a significant feature in this block is that the overlap between the positions of $\vec{1}$
in the $j$th and $(j+1)$th lines halves. More specifically to say, compared to the position of $\vec{1}$ in the $j$th line,
only the left (right) half in the same position keeps being $1$ in the $(j+1)$th line if $j$ is odd (even). For example, in
the first line ($j=1$) all entries are $1$, and hence the length of $\vec{1}$ is $2^{n-2}$. In the second line ($j=2$), only
the left half keeps being one, and the length of $\vec{1}$ becomes $2^{n-3}$. Similarly, in the third line ($j=3$) only the
right half keeps being $1$ compared to the position of $\vec{1}$ in the second line. Utilizing this feature, one can find
that when $n$ is even, the $\frac{1}{3}(2^n-1)$th and $\frac{1}{3}(2^n+2)$th entries keep being $1$ in the $(n-2)$th line.
Notice that the entry number here starts from the beginning of all diagonal entries of $-\sigma^z_j\sigma^z_{j+1}$, not the
beginning of the second block. And in the $(n-1)$th line, the $\frac{1}{3}(2^n+2)$th entry is $1$. In the case of open boundary
condition, this is the last line and the position is located. In the case of the periodic boundary condition, one more line of
$-\sigma^z_{n}\sigma^z_{1}$ needs to be considered. Luckily, this position of $-\sigma^z_{n}\sigma^z_{1}$ is also $1$ when $n$
is even. Therefore, the maximum energy is at the $\frac{1}{3}(2^n+2)$th entry under both boundary conditions. Due to the
symmetry, the $\frac{1}{3}(2^{n+1}+1)$th entry, which is in the third block, is also maximum.

Now we locate the values of $\frac{1}{3}(2^n+2)$th and $\frac{1}{3}(2^{n+1}+1)$th entries in $\sigma^z_j$, which is
irrelevant to the boundary condition. The block of entries in $\sigma^z_j$ with respect to the second block in
Fig.~\ref{fig:entrysearch}(b) is given in Fig.~\ref{fig:entrysearch}(c). As shown in this diagram, the $\frac{1}{3}(2^n+2)$th
entry is $1$ for an odd $j$ and $-1$ for an even $j$, namely, it is $(-1)^{j+1}$. Similarly, one can find that the
$\frac{1}{3}(2^{n+1}+1)$th entry is $(-1)^{j}$. Hence, $D_j=(-1)^{j+1}\sigma_z$. In the meantime, both $\frac{1}{3}(2^n+2)$th
and $\frac{1}{3}(2^{n+1}+1)$th entries are zero in $J_z$ when $n$ is even, which means $G=0$.

In summary, we have found that $A=n\sigma_z/2$, $G=0$, $C_j=\sigma_z$, and $D_j=(-1)^{j+1}\sigma_z$. Utilizing these
expressions, Eqs.~(\ref{eq:apx_xig}) and (\ref{eq:apx_xil}) can be further written into
\begin{equation}
\partial_t\xi_t=\left[i(E_{\max}-E_{\min})-\frac{n^2\gamma_g}{8}\right]\xi_t,
\label{eq:apx_xigtp1}
\end{equation}
and
\begin{equation}
\partial_t \xi_t=i(E_{\max}-E_{\min})\xi_t+\sum_j\gamma_{l,j}\left[(-1)^{j+1}\sigma_z\xi_t\sigma_z-\xi_t\right].
\label{eq:apx_xiltp1}
\end{equation}
Equation~(\ref{eq:apx_xigtp1}) can be easily solved as
\begin{equation}
\xi_t=e^{\left[i(E_{\max}-E_{\min})-\frac{n^2\gamma_g}{8}\right]}\xi,
\end{equation}
and Eq.~(\ref{eq:apx_xiltp1}) can be solved as
\begin{align*}
[\xi_t]_{00(11)} =& e^{i(E_{\max}-E_{\min})-\sum^n_{j=1}\gamma_{l,j}\left[1+(-1)^j\right]}[\xi]_{00(11)}, \\
[\xi_t]_{01(10)} =& e^{i(E_{\max}-E_{\min})-\sum^n_{j=1}\gamma_{l,j}\left[1-(-1)^j\right]}[\xi]_{01(10)}.
\end{align*}
Here $[\cdot]_{ab}$ represents the $ab$th entry ($a,b=0,1$).

Next we calculate $\cos(\theta(t))$. Notice that Eq.~(\ref{eq:apx_isingcos}) can be expressed by
\begin{equation}
\cos(\theta(t))=\frac{\mathrm{Re}\left(\mathrm{Tr}(\xi\xi^{\dagger}_t)\right)}
{\sqrt{\mathrm{Re}\left(\mathrm{Tr}(\xi\xi^{\dagger})\right)
\mathrm{Re}\left(\mathrm{Tr}(\xi_t\xi^{\dagger}_t)\right)}},
\label{eq:apx_costp1}
\end{equation}
where $\mathrm{Re}(\cdot)$ represents the real part. In the case of global dephasing [Eq.~(\ref{eq:apx_xigtp1})],
the expression above reduces to
\begin{equation}
\cos(\theta(t))=\cos\left((E_{\max}-E_{\min})t\right),
\end{equation}
which is irrelevant to the decay rate $\gamma$. Hence, the evolution time to reach the target for the optimal states
in Eq.~(\ref{eq:apx_optstate1}) is indeed robust to the global dephasing in both periodic and open boundary conditions,
indicating that their difference is also robust.

%========================================= Figure =========================================
\begin{figure}[tp]
\centering\includegraphics[width=8.5cm]{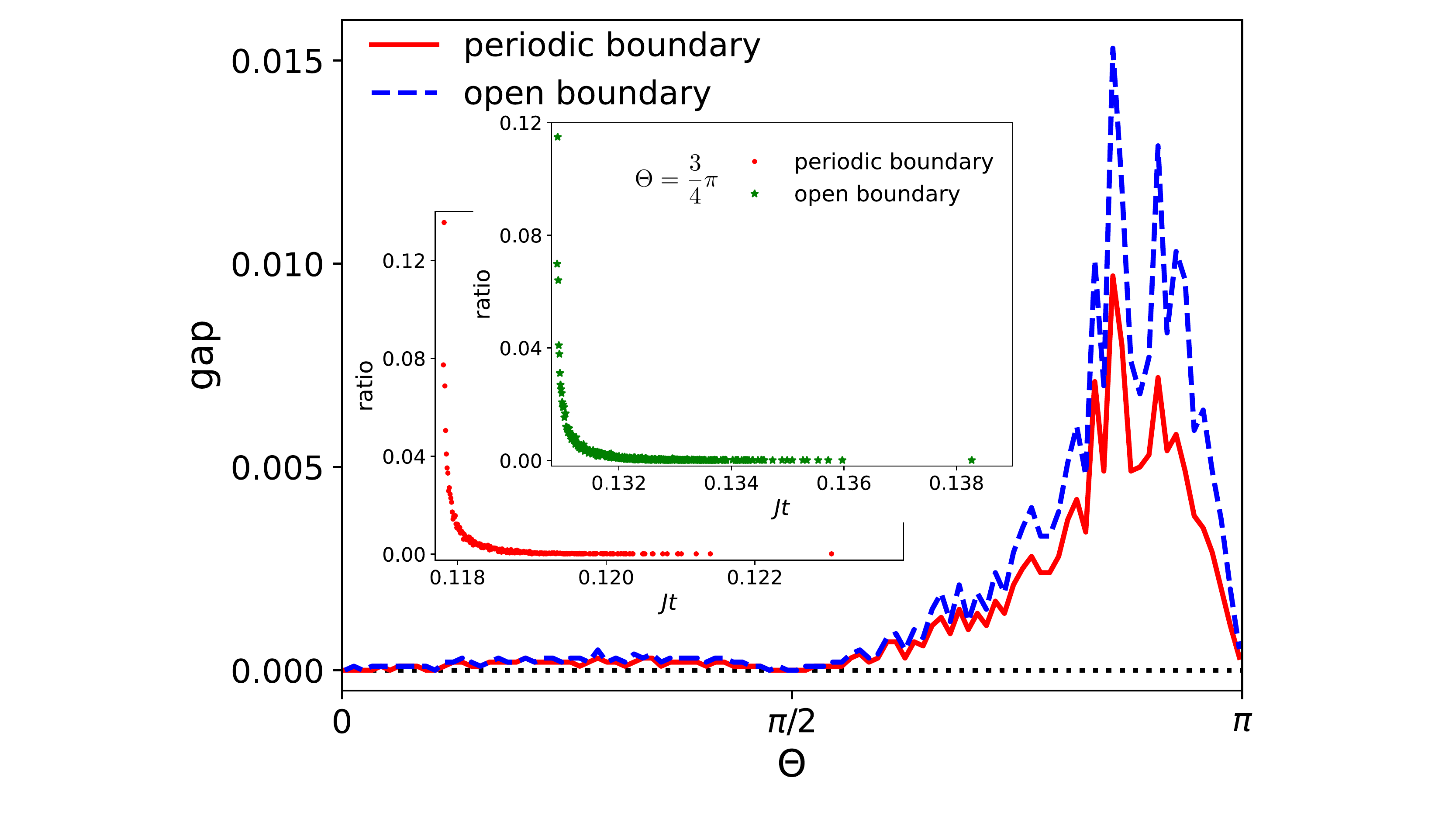}
\caption{The variety of the gap between the maximum and minimum values of the evolution
time to reach the target $\Theta$ among $100$ random states with random values of
$\{\gamma_{l,j}\}\in (0,1)$. The insets present the ratios of $10000$ states at different
evolution time to reach the target $\Theta=3\pi/4$ for periodic (red dots) and open (green
pentagrams) boundary conditions. $n=10$ in all plots.}
\label{fig:robustness_theta}
\end{figure}
%==========================================================================================

In the case of local dephasing, Eq.~(\ref{eq:apx_costp1}) can be expressed by
\begin{eqnarray*}
& & \cos(\theta(t)) \nonumber \\
&=& \cos\left((E_{\max}-E_{\min})t\right)\frac{\varsigma_1+\varsigma_2 e^{-2t\gamma_{\mathrm{all}}}}
{\sqrt{\varsigma_1+\varsigma_2}\sqrt{\varsigma_1+\varsigma_2 e^{-4t\gamma_{\mathrm{all}}}}},
\end{eqnarray*}
where $\varsigma_1=|[\xi_t]_{00}|^2+|[\xi_t]_{11}|^2$, $\varsigma_2=|[\xi_t]_{01}|^2+|[\xi_t]_{10}|^2$, and
$\gamma_{\mathrm{all}}=\sum^n_{j=1}\gamma_{l,j} (-1)^j$. If the values of all decay rates $\{\gamma_{l,j}\}$ are very
close, for example $\gamma_{l,j}\approx\gamma$ for any $j$, then $\gamma_{\mathrm{all}}\approx 0$ and $\cos(\theta(t))$
still approximates to $\cos\left((E_{\max}-E_{\min})t\right)$, which is also irrelevant to the decay rates, and thus in
this case the evolution time, as well as the time difference, are also robust to the local dephasing. In the case that
the values of $\{\gamma_{l,j}\}$ are not close, Eq.~(\ref{eq:apx_costp1}) is indeed dependent on the decay rates.
However, since $\varsigma_1+\varsigma_2 e^{-2t\gamma_{\mathrm{all}}}$ is always positive at finite time, the evolution
time is still irrelevant to $\gamma_{\mathrm{all}}$ for the target $\Theta=\pi/2$ and hence robust to the local
dephasing. For a general target, we have tested $100$ random states in Eq.~(\ref{eq:apx_optstate1}) with random values
of $\{\gamma_{l,j}\}\in (0,1)$ for each target in the case of $n=10$, and the gap between the maximum and minimum values
of the evolution time for these $100$ states are given in Fig.~\ref{fig:robustness_theta}. It can be seen that the
robustness is quite good when the target is no larger than $\pi/2$, and it is indeed compromised when $\Theta$ is larger
than $\pi/2$. Even for those targets with large gaps, the evolution time for different states could concentrate on some
specific values, namely, the distribution of states in the gap has a sharp peak. For example, the insets of
Fig.~\ref{fig:robustness_theta} show the distributions of $10000$ states for periodic (red dots) and open (green pentagrams)
boundary conditions in the case of $\Theta=3\pi/4$. It can be seen that the distributions for both periodic and open
boundary conditions have a sharp peak at the minimum values, indicating that the evolution time is still relatively
robust for most states.

\section{Learning the OQSL in Landau-Zener model}
\label{sec:apx_LZ}

\subsection{Verification of the validity of CRC methodology}

%============================================ Figure =============================================================
\begin{figure*}[tp]
\centering\includegraphics[width=17.5cm]{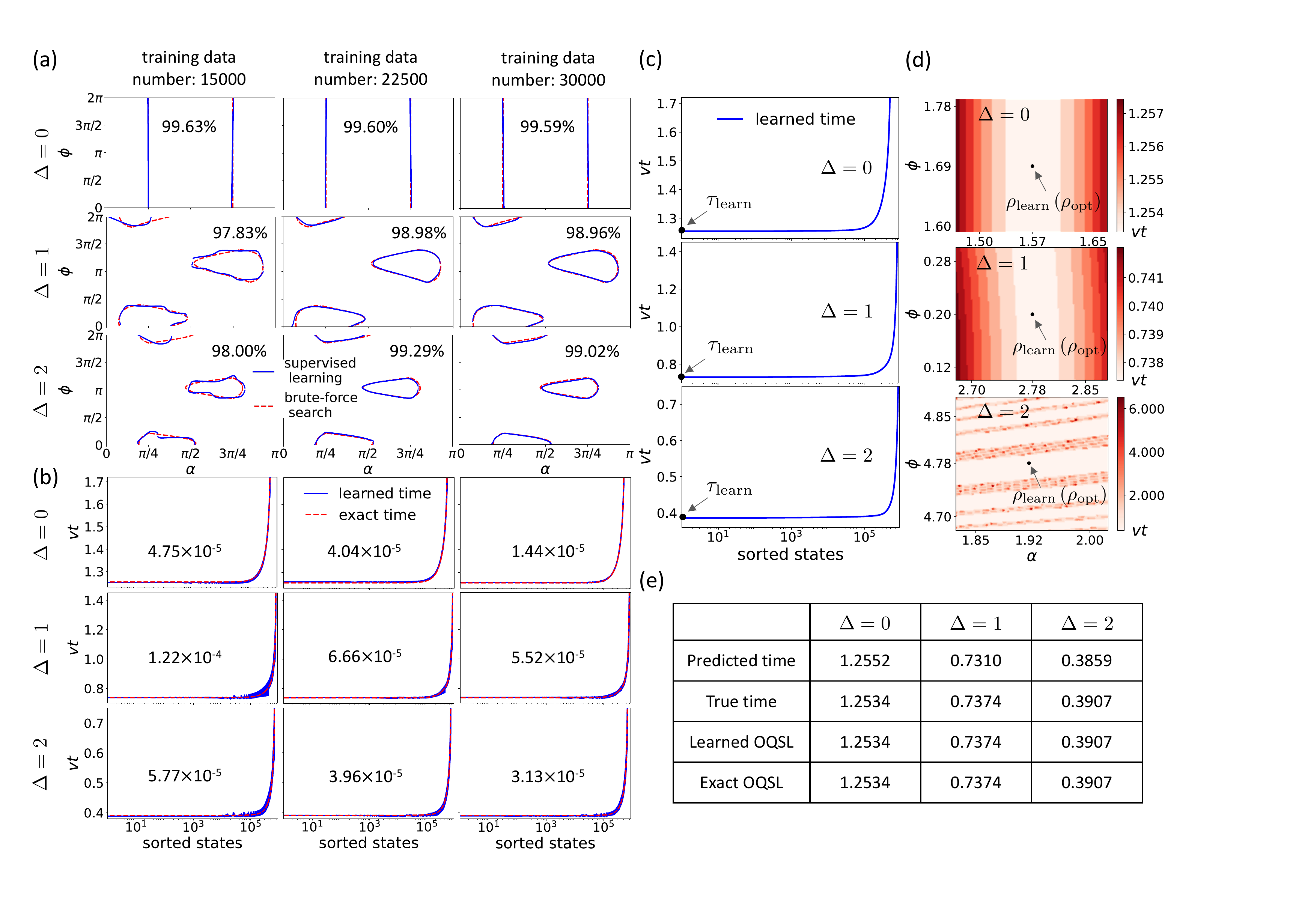}
\caption{(a) Comparison between the set $\mathcal{S}$ (brute-force search) and $\mathcal{S}_{\mathrm{learn}}$
(learning) with different values of $\Delta$ and different training data number. The first, second, and third
rows represent the results for $\Delta=0$, $\Delta=1$, and $\Delta=2$, respectively. The first, second, and
third columns represent the results for 15000, 22500, and 30000 training data, respectively. The solid blue
(dashed red) lines represent the boundaries between $\mathcal{S}$ ($\mathcal{S}_{\mathrm{learn}}$) and its
complementary set. The percentage numbers in the plots are the scores of the learning. (b) Comparison of the
evolution time to reach the target obtained from the regression process (solid blue lines) and the exact time
obtained from the brute-force search (dashed red lines). Here the input states in the regression are the ones
in $\mathcal{S}$. The numbers in the plots are the mean square errors of learning. (c) The practical performance
of the regression process where the input states are those in $\mathcal{S}_{\mathrm{learn}}$. (d) Results of
the calibration process. The region for calibration is taken as $[\alpha_{\mathrm{learn}}-0.1,\alpha_{\mathrm{learn}}+0.1]$
and $[\phi_{\mathrm{learn}}-0.1,\phi_{\mathrm{learn}}+0.1]$. The black dots represent $(\alpha_{\mathrm{learn}},
\phi_{\mathrm{learn}})$, the "optimal" points obtained in the regression. (e) Table of the predicted time
($\tau_{\mathrm{learn}}$) obtained in the practical regression process, the corresponding true time, and finally
learned OQSL after the calibration process for different values of $\Delta$. The exact OQSL is obtained via
brute-force search. In all plots $v$ is set to be 1 and the target angle $\Theta=\pi/2$.}
\label{fig:apx_LZ}
\end{figure*}
%=================================================================================================================

Here we present the process of learning the OQSL in the Landau-Zener model and show the validity of the CRC methodology.
The Hamiltonian of this model is
\begin{equation}
H=\Delta \sigma_{x}+vt\sigma_z,
\end{equation}
where $\Delta$ and $v$ are two time-independent parameters. $\sigma_x$ and $\sigma_z$ are the Pauli matrices.
The OQSL in this model has been thoroughly discussed in Ref.~\cite{Shao2020}, in which the set $\mathcal{S}$
is obtained via the brute-force search among around one million pure states. The reason why only pure states
are considered here is due to the fact that unitary evolution does not affect the purity and in the Bloch
representation all states in the same direction can/cannot reach the target simultaneously. The dynamics is
solved via QuTiP~\cite{Johansson2012,Johansson2013}. The full evolution is truncated at $vt=10$, namely, the
state is treated to not be in $\mathcal{S}$ if it cannot reach the target within the truncated time. The Bloch
vector of the initial state is parameterized by $\vec{r}=(\sin\alpha\cos\phi,\sin\alpha\sin\phi,\cos\alpha)^{\mathrm{T}}$
with $\alpha\in[0,\pi]$ and $\phi\in[0,2\pi)$.

In the step of classification, a multilayer neural network with two inputs ($\alpha$ and $\phi$) and one
output (1 or 0) is created with a hyperbolic tangent function as the activation function. The output result
$1$/$0$ represents that the input initial state can/cannot realize the given target, respectively. Supervised
learning is performed via scikit-learn~\cite{Pedregosa2011}. The network contains five to six hidden layers each
with about 200 to 250 neurons. The Cross-Entropy loss function~\cite{Baum1988} is used as the loss function, and
Adam~\cite{Kingma2014} is applied in the updates of the network. The test set contains all the initial states
(around one million states) used in the brute-force search. The performance of training for different values of
$\Delta$ are given in Fig.~\ref{fig:apx_LZ}(a). The first, second, and third rows represent the learned $\mathcal{S}$
for $\Delta=0$, $\Delta=1$, and $\Delta=2$ (in the units of $\sqrt{v}$). The solid blue and dashed red lines represent
the boundaries between $\mathcal{S}$ and its complementary set obtained via supervised learning and brute-force search.
Different numbers of the training set, including 15000, 22500, and 30000, have also been tested and compared, as shown
in the first (15000), second (22500), and third (30000) columns in Fig.~\ref{fig:apx_LZ}(a). The percentage numbers in
the plots are the scores of learning, i.e., the correctness of the network's output. It can be seen that the performance
of 15000 training data is better than the others in the case of $\Delta=0$, and 22500 training data present the best
performance in the cases of $\Delta=1$ and $\Delta=2$. One should notice that all the parameters of the network are
manually tuned case by case, and the slight difference in the performance may not be fully due to the difference
in the training data number. In the case of 22500 training data, the correctness is around $99\%$, indicating that
about $0.99$ million states are correctly classified into $\mathcal{S}$ and its complementary set. Therefore, the
neural network indeed works for the classification in this example.

The second step is the regression process, in which basically the same neural network is created but with rectified
linear unit function as the activation function. The loss function is taken as the square error loss
function~\cite{Lehmann1998}. The training data are sorted by the evolution time to reach the target from smallest to
largest. Similar to the classification process, all the states in $\mathcal{S}$ are used to test the performance of
the network. Notice that $\mathcal{S}$ here is the exact reachable state set obtained via the brute-force search
since we need to check the validity of the network. The performance of regression is presented for different values
of $\Delta$ and training data number in Fig.~\ref{fig:apx_LZ}(b). All the plots in this figure are semi-logarithmic
($x$ axis). The first, second, and third rows represent the results for $\Delta=0$, $\Delta=1$, and $\Delta=2$. The
first, second, and third columns represent the results for 15000, 22500, and 30000 training data. The number in the
plots are the mean square errors of learning, i.e., $\frac{1}{m}\sum^{m}_{i=1}\big[t^{(i)}_{\mathrm{pre}}-
t^{(i)}_{\mathrm{ext}}\big]^2$. Here $t^{(i)}_{\mathrm{pre}}$ and $t^{(i)}_{\mathrm{ext}}$ are the predicted time
obtained via learning and exact time obtained via brute-force search for the $i$th state. The order of states in the
figure is sorted by the evolution time obtained in the brute-force search from smallest to largest, and the learned
time is plotted using the same order of states. Notice that these states are not exactly the same for different values
of $\Delta$ due to the dependence of $\mathcal{S}$ on $\Delta$. It can be seen that the performance of learning
(solid blue lines) is good for all values of $\Delta$, especially when the training data number is 22500 and 30000.
Basically the mean square errors of learning in these two cases for all values of $\Delta$ are in the scale of $10^{-5}$.
Hence, the network also works for the regression in this example.

As a matter of fact, in practice the reachable state set used in the regression process is the one obtained in the
classification process (denoted by $\mathcal{S}_{\mathrm{learn}}$). Hence, although it is reasonable to use the
true $\mathcal{S}$ to check the validity of the regression, $\mathcal{S}_{\mathrm{learn}}$ has to be applied to
test if the OQSL obtained from CRC methodology is reasonable. The performance of regression with respect to
$\mathcal{S}_{\mathrm{learn}}$ for 22500 training data is given in Fig.~\ref{fig:apx_LZ}(c) for different values
of $\Delta$. The results for the other two training data numbers are not shown here due to their similarity. Since
the training set chosen in Fig.~\ref{fig:apx_LZ}(b) is also a subset of $\mathcal{S}_{\mathrm{learn}}$, we can
directly use it as the training set in this case and the trained network is then the same. The states in the plots
are sorted by the evolution time to reach the target from smallest to largest. As shown in this figure, the trend
of learned time basically coincides with the exact time in Fig.~\ref{fig:apx_LZ}(b). One should notice that in fact
these two lines cannot be compared directly as the states are not exactly the same. Utilizing the result of the
regression, the "optimal" state $\rho_{\mathrm{learn}}$ and corresponding predicted time $\tau_{\mathrm{learn}}$ can
be located. The rigorous evolution time of $\rho_{\mathrm{learn}}$ to reach the target (true time) is given in the
table in Fig.~\ref{fig:apx_LZ}(e). It can be seen that the predicted time $\tau_{\mathrm{learn}}$ is very close to
the true time for all values of $\Delta$. The errors in all cases are on the scale of $10^{-3}$, indicating that the
regression process works well in this example. Furthermore, the true time of $\rho_{\mathrm{learn}}$ coincides
with the exact OQSL obtained via brute-force search, which means $\rho_{\mathrm{learn}}$ is indeed an optimal state
in this case.

The last process is calibration. The core of this process is to calculate the rigorous dynamics of the states
around $\rho_{\mathrm{learn}}$ and find the exact minimum time in this region. This process guarantees the finally
obtained time is the rigorous minimum time in this region. In this example, the values of $(\alpha,\phi)$ for the
"optimal" states [denoted by $(\alpha_{\mathrm{learn}},\phi_{\mathrm{learn}})$] in the cases of $\Delta=0$, $\Delta=1$,
and $\Delta=2$ are $(1.57,1.69)$, $(2,78,0.20)$, and $(1.92,4.78)$, respectively [black dots in Fig.~\ref{fig:apx_LZ}(d)].
The region to perform the calibration is $[\alpha_{\mathrm{learn}}-0.1,\alpha_{\mathrm{learn}}+0.1]$ and
$[\phi_{\mathrm{learn}}-0.1,\phi_{\mathrm{learn}}+0.1]$. The rigorous dynamics of about 10000 states in this region
are calculated. The results are given in Fig.~\ref{fig:apx_LZ}(d) and the corresponding minimum time (learned OQSL)
is given in the table in Fig.~\ref{fig:apx_LZ}(e). The consistency between the learned OQSL and the exact OQSL proves
that the final result is indeed the exact OQSL in this example. The validity of the CRC methodology is then confirmed.

\subsection{Learning the OQSL in the controlled system}

Next, we apply the CRC methodology to search the OQSL in the controlled Landau-Zener model. The full Hamiltonian of
this model reads
\begin{equation}
H=\Delta \sigma_x + vt\sigma_z + \vec{u}\cdot \vec{\sigma},
\label{eq:apx_LZctrl}
\end{equation}
where $\vec{\sigma}=(\sigma_x, \sigma_y, \sigma_z)$ is the vector of Pauli matrices, and $\vec{u}=(u_x, u_y, u_z)$
is the vector of control amplitudes.

%============================================= Algorithm =============================================
\begin{algorithm}[tp]
%\SetAlgoNoLine
\SetArgSty{<texttt>}
\caption{auto-GRAPE}
Initialize the control amplitude $u_k(t)$ for all $t$ and $k$; \\
\For {episode=1, $M$}{
Receive initial state $\rho_0$; \\
\For {$t=1, T$}{
Evolve with the control $\rho_t=e^{\Delta t \mathcal{L}_t}\rho_{t-1}$; \\
Calculate $f_t=\frac{N\mathrm{Tr}(\rho_0\rho_t)-1}{\sqrt{\left[N\mathrm{Tr}(\rho^2_0)-1\right]
\left[N\mathrm{Tr}(\rho^2_t)-1\right]}}$ and save it; \\
}
Calculate the objective function $f=\sum_t f_t$. \\
Calculate the gradient $\frac{\delta f}{\delta u_k(t)}$ with the automatic differentiation method
for all $t$ and $k$.\\
{\For {$t=1, T$}{
\For {$k=1, K$}{
Update control $u_k(t)\!\leftarrow\! u_k(t)\!+\!\epsilon\frac{\delta f}{\delta u_k(t)}$.
}}}
}
Save the controls $\{u_k\}$.
\label{algorithm:autogrape}
\end{algorithm}
%======================================================================================================

We first discuss the generation of controls for a specific initial state to reach the target at the minimum time. The
controls are generated via the auto-GRAPE~\cite{Zhang2022} with the objective function
\begin{equation}
f=\int^{T}_0\cos\left(\theta (t)\right)\mathrm{d}t,
\end{equation}
where $T$ is a reasonably long time (truncated time in our calculation), and $\theta(t)$ is the angle between the
Bloch vectors of initial state $\rho_0$ and its evolved state $\rho_t$, which satisfies the equation $\partial_t
\rho_t=\mathcal{L}_t\rho_t$ with $\mathcal{L}_t$ a time-dependent superoperator. Notice that in the Bloch representation
the density matrix can be expressed by Eq.~(\ref{eq:apx_dmBloch}). Then $\cos(\theta(t))$ can be calculated by
\begin{equation}
\cos(\theta(t))=\frac{N\mathrm{Tr}(\rho_0\rho_t)-1}{\sqrt{\left[N\mathrm{Tr}(\rho^2_0)-1\right]
\left[N\mathrm{Tr}(\rho^2_t)-1\right]}}.
\end{equation}
In this case, the dynamics is unitary and only pure states need to be calculated, then $\cos(\theta(t))$ reduces to
$2\mathrm{Tr}(\rho_0\rho_t)-1$. In the numerical calculation, the evolution time is usually discretized into many
equally spaced time points ($\{t_i\}$), and thus we can use the discrete form
\begin{equation}
f=\sum_{i}\cos(\theta(t_i))
\label{eq:apx_obj}
\end{equation}
as the objective function instead. The time interval here is neglected since it does not affect the final performance. 
In the numerical calculation, the difference between the discretization error of the integration and the value 
of the objective function is at the scaling of $10^{-8}$ and thus this error would not cause any significant effect on 
the final result. 

%================================== Figure ===================================
\begin{figure}[tp]
\centering\includegraphics[width=8.5cm]{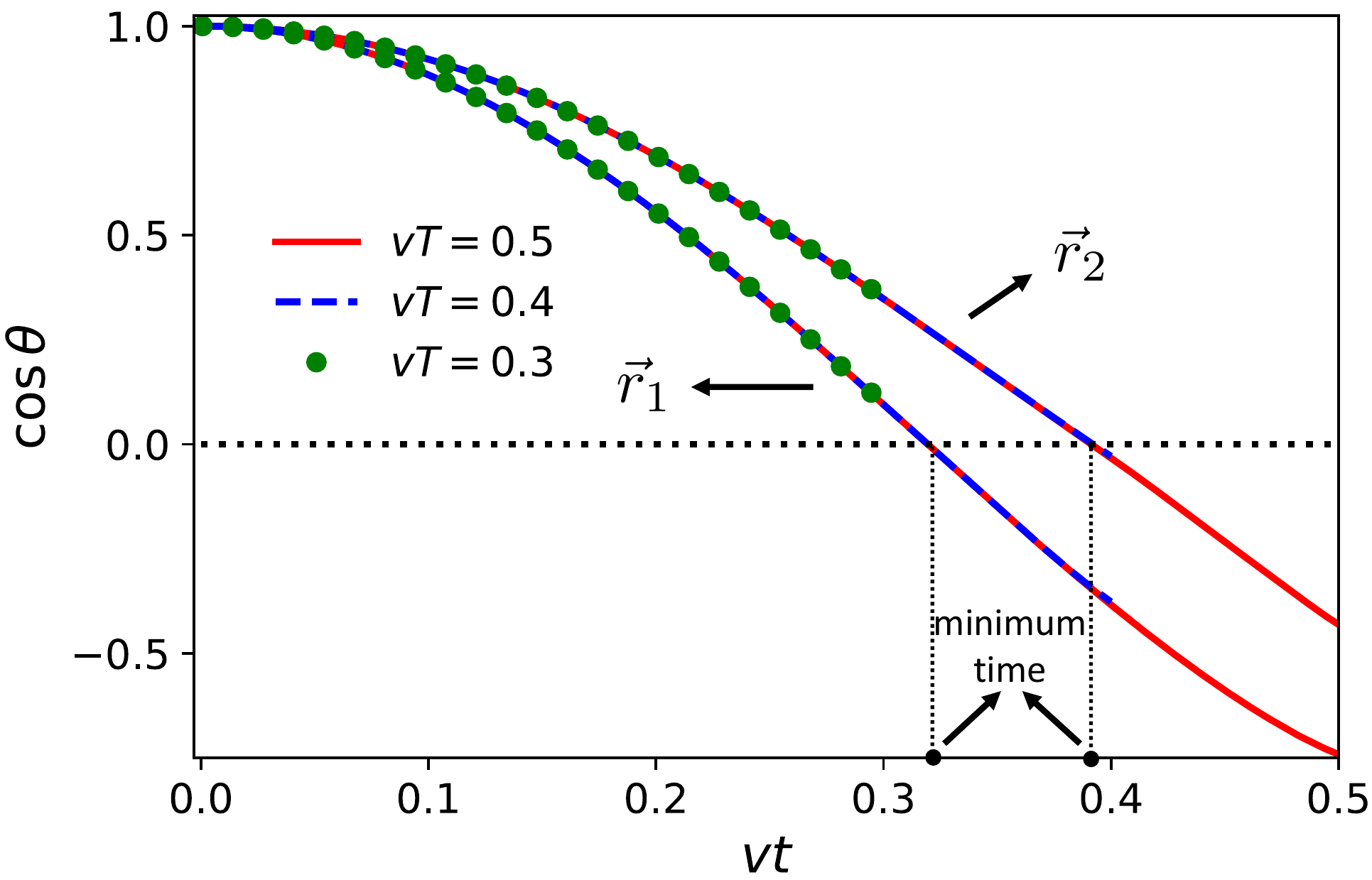}
\caption{Performance of controls for two randomly generated initial
states $\vec{r}_1$ and $\vec{r}_2$ with different values of $T$ (in the unit
of $v$), including $T=0.3$ (green dots), $T=0.4$ (dashed blue lines), and $T=0.5$
(solid red lines).}
\label{fig:apx_LZobj}
\end{figure}
%=============================================================================

Auto-GRAPE is a gradient-based algorithm where the gradient is evaluated via automatic differentiation~\cite{Zhang2022}.
In Ref.~\cite{Zhang2022} the quantum metrological quantities like quantum Fisher information are taken as the objective
function, here in this paper we take Eq.~(\ref{eq:apx_obj}) as the objective function. The corresponding pseudocode is
given in Algorithm~\ref{algorithm:autogrape}. In one episode, the initial state is evolved to time $T$ and the objective
function is calculated. Then the gradients $\delta f/\delta u_k(t)$ for all $t$ and $k$ are evaluated via automatic
differentiation, which is realized with the Julia package Zygote~\cite{Innes2018}. At last, all the control amplitudes
are updated simultaneously according to the evaluation of gradients. In practice, Adam~\cite{Kingma2014} could be applied
to further improve efficiency.

%======================================== Figure ========================================
\begin{figure*}[tp]
\centering\includegraphics[width=17.8cm]{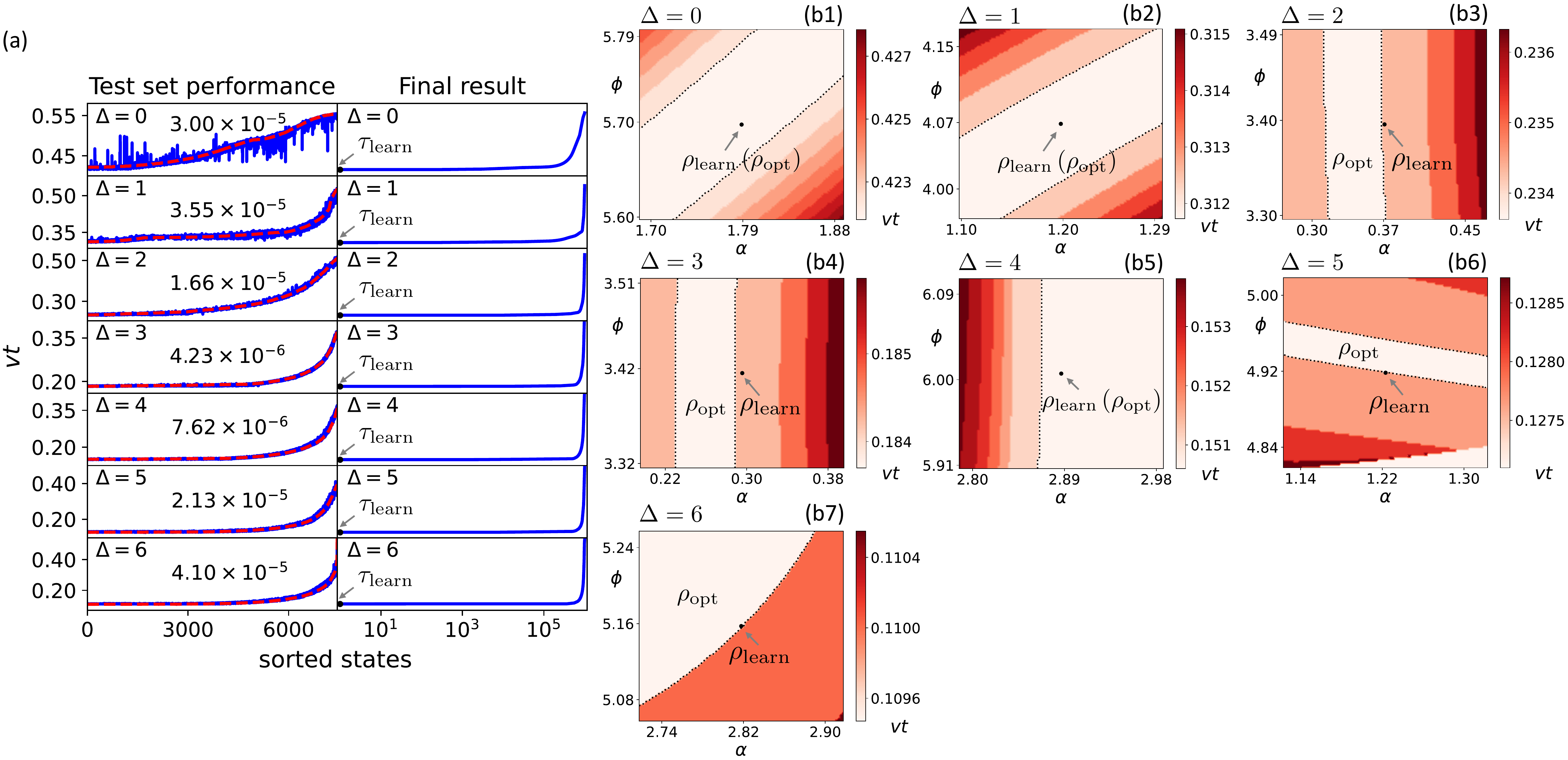}
\caption{CRC methodology for controlled noiseless dynamics. (a) The left column: The test
set performance of the regression process for $\Delta=0,1,2,3,4,5,6$ (top to bottom). The
right column: The result of regression for $\Delta=0,1,2,3,4,5,6$ (top to bottom). The
solid blue and dashed red lines represent the learned time and exact time obtained from
regression and rigorous dynamics, respectively. The numbers in the left column are the
mean square errors of learning. The $x$ axes in both columns are in the logarithmic scales.
[(b1)-(b7)] Results of calibration for $\Delta=0,1,2,3,4,5,6$. The regime for calibration
is $[\alpha_{\mathrm{learn}}-0.1,\alpha_{\mathrm{learn}}+0.1]$ and $[\phi_{\mathrm{learn}}-0.1,
\phi_{\mathrm{learn}}+0.1]$. The target $\Theta$ is taken as $\pi/2$ in all plots.}
\label{fig:LZctrl}
\end{figure*}
%========================================================================================

To test the validity of the objective function [Eq.~(\ref{eq:apx_obj})], the performance of corresponding controls are
demonstrated in Fig.~\ref{fig:apx_LZobj} for two randomly generated initial states, of which the Bloch vectors are
$(0.22,0.20,-0.96)^{\mathrm{T}}:=\vec{r}_1$ and $(0.95,-0.15,-0.29)^{\mathrm{T}}:=\vec{r}_2$. Three different values
of $T$ (in the unit of $v$), including $T=0.3$ (green dots), $T=0.4$ (dashed blue lines), and $T=0.5$ (solid red lines) are
tested. As shown in the figure, the optimal controls for $T=0.4$ and $0.5$ can let the states reach the target angle
(dotted black line) at the same time, confirming that this found time (black dots) is indeed minimum. In the meantime, if
$T$ is smaller than the minimum time, for example $T=0.3$, the states cannot reach the target angle during the entire
evolution, which also corroborates that the found time is minimum as the states cannot reach the target before this time.
Hence, the validity of the objective function and corresponding controls are confirmed. Moreover, the consistency of
performance for $T=0.4$ and $T=0.5$ shows that the choice of $T$ does not affect the result of minimum time as long
as it is larger than the minimum time.

%======================================== Figure ========================================
\begin{figure*}[tp]
\centering\includegraphics[width=17.8cm]{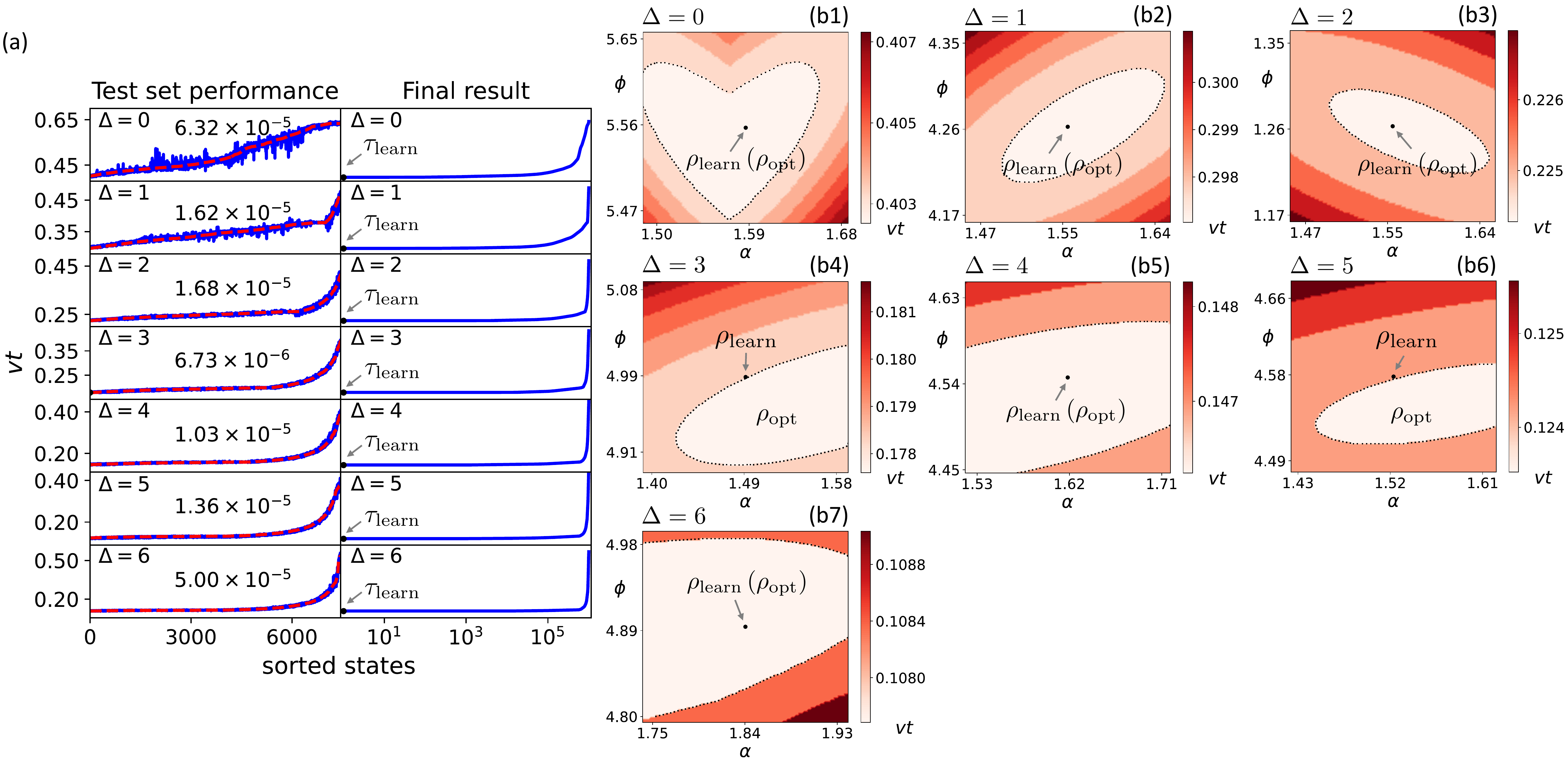}
\caption{CRC methodology for controlled noisy dynamics. (a) The left column: The test
set performance of the regression process for $\Delta=0,1,2,3,4,5,6$ (top to bottom). The
right column: The result of regression for $\Delta=0,1,2,3,4,5,6$ (top to bottom). The
solid blue and dashed red lines represent the learned time and exact time obtained from
regression and rigorous dynamics, respectively. The numbers in the left column are the
mean square errors of learning. The $x$ axes in both columns are in the logarithmic scales.
[(b1)-(b7)] Results of calibration for $\Delta=0,1,2,3,4,5,6$. The regime for calibration
is $[\alpha_{\mathrm{learn}}-0.1,\alpha_{\mathrm{learn}}+0.1]$ and $[\phi_{\mathrm{learn}}-0.1,
\phi_{\mathrm{learn}}+0.1]$. The target $\Theta$ is taken as $\pi/2$ in all plots.}
\label{fig:LZctrl_noise}
\end{figure*}
%=========================================================================================

Next, we perform the CRC methodology for $\Delta=0,1,2,3,4,5,6$ (in the units of $\sqrt{v}$) in both noiseless and noisy cases.
In the noiseless case, the result of classification shows that all states in the state space can fulfill the target. This phenomenon
is reasonable in physics due to the full controllability of $\vec{u}\cdot\vec{\sigma}$, which means the controls can realize the
rotation of a state from any angle. Thus, any state can fulfill the target in finite time under this control Hamiltonian even without
the free Hamiltonian $\Delta\sigma_x+vt\sigma_z$.

In the step of regression, the data number of training and test sets are 22500 and 7500. Similar to the noncontrolled case,
the mean square errors between the learned time (solid blue lines) and exact time (dashed red lines) in the test set are still
on the scales of $10^{-5}$ and $10^{-6}$ for all values of $\Delta$, as shown in the left column in Fig.~\ref{fig:LZctrl}(a).
Utilizing this learned regression network, about one million states are input and corresponding learned time (solid blue lines)
is given in the right column in Fig.~\ref{fig:LZctrl}(a). The minimum time $\tau_{\mathrm{learn}}$ for $\Delta=0,1,2,3,4,5,6$
are $0.4154$, $0.3080$, $0.2315$, $0.1830$, $0.1486$, $0.1261$, and $0.1113$, respectively.

In the last step, the region for calibration is still taken as $[\alpha_{\mathrm{learn}}\!-0.1,\alpha_{\mathrm{learn}}+0.1]$
and $[\phi_{\mathrm{learn}}-0.1,\phi_{\mathrm{learn}}+0.1]$. The results of calibration are given in
Figs.~\ref{fig:LZctrl}(b1)-\ref{fig:LZctrl}(b7). As shown in the plots, the optimal state $\rho_{\mathrm{opt}}$ coincides with
$\rho_{\mathrm{learn}}$ in the cases of $\Delta=0,1,4$. However, the position of $\rho_{\mathrm{opt}}$ slightly moves away
from $\rho_{\mathrm{learn}}$ in other cases, which proves the necessity of the step of calibration.

In the noisy case, the dephasing is invoked and the dynamics of the density matrix is governed by the master equation
\begin{equation}
\partial_t\rho_t=-i[H,\rho_t]+\gamma(\sigma_z\rho_t\sigma_z-\rho_t),
\end{equation}
where the Hamiltonian is in Eq.~(\ref{eq:apx_LZctrl}) and $\gamma$ is the decay rate, which is taken as $0.5\sqrt{v}$ in
the following calculation. The CRC methodology has been applied in this case for $\Delta=0,1,2,3,4,5,6$ (in the units of
$\sqrt{v}$). The result of the classification here is the same as that in the noiseless case, i.e., all states can fulfill
the target under control. The results of regression and calibration are given in Fig.~\ref{fig:LZctrl_noise}. Similar to
the noiseless case, the mean square errors of regression are still on the scales of $10^{-5}$ and $10^{-6}$, as shown in
Fig.~\ref{fig:LZctrl_noise}(a). The minimum time $\tau_{\mathrm{learn}}$ for $\Delta=0,1,2,3,4,5,6$ are $0.3961$, $0.2968$,
$0.2242$, $0.1783$, $0.1440$, $0.1196$, and $0.1063$, respectively. In the calibration, the region for calibration is still
taken as $[\alpha_{\mathrm{learn}}\!-\!0.1,\alpha_{\mathrm{learn}}\!+\!0.1]$ and $[\phi_{\mathrm{learn}}\!-\!0.1,
\phi_{\mathrm{learn}}\!+\!0.1]$, as shown in Figs.~\ref{fig:LZctrl_noise}(b1)-\ref{fig:LZctrl_noise}(b7) for different values
of $\Delta$. The results show that $\rho_{\mathrm{opt}}$ are either the same with $\rho_{\mathrm{learn}}$, or very close to it,
indicating that both regression and calibration processes are effective in this case.

In this example, the time costs for the generation of training sets and the training of the neural networks are all less than half an 
hour on a regular personal computer, and that for the calibration is around 5 minutes for a value of $\Delta$. Hence, the CRC 
methodology can be easily applied to the few-body systems without any extra requirement on the computational setup. 

\section{The OQSL for transverse Ising model}

%======================================== Figure ========================================
\begin{figure}[tp]
\centering\includegraphics[width=8.5cm]{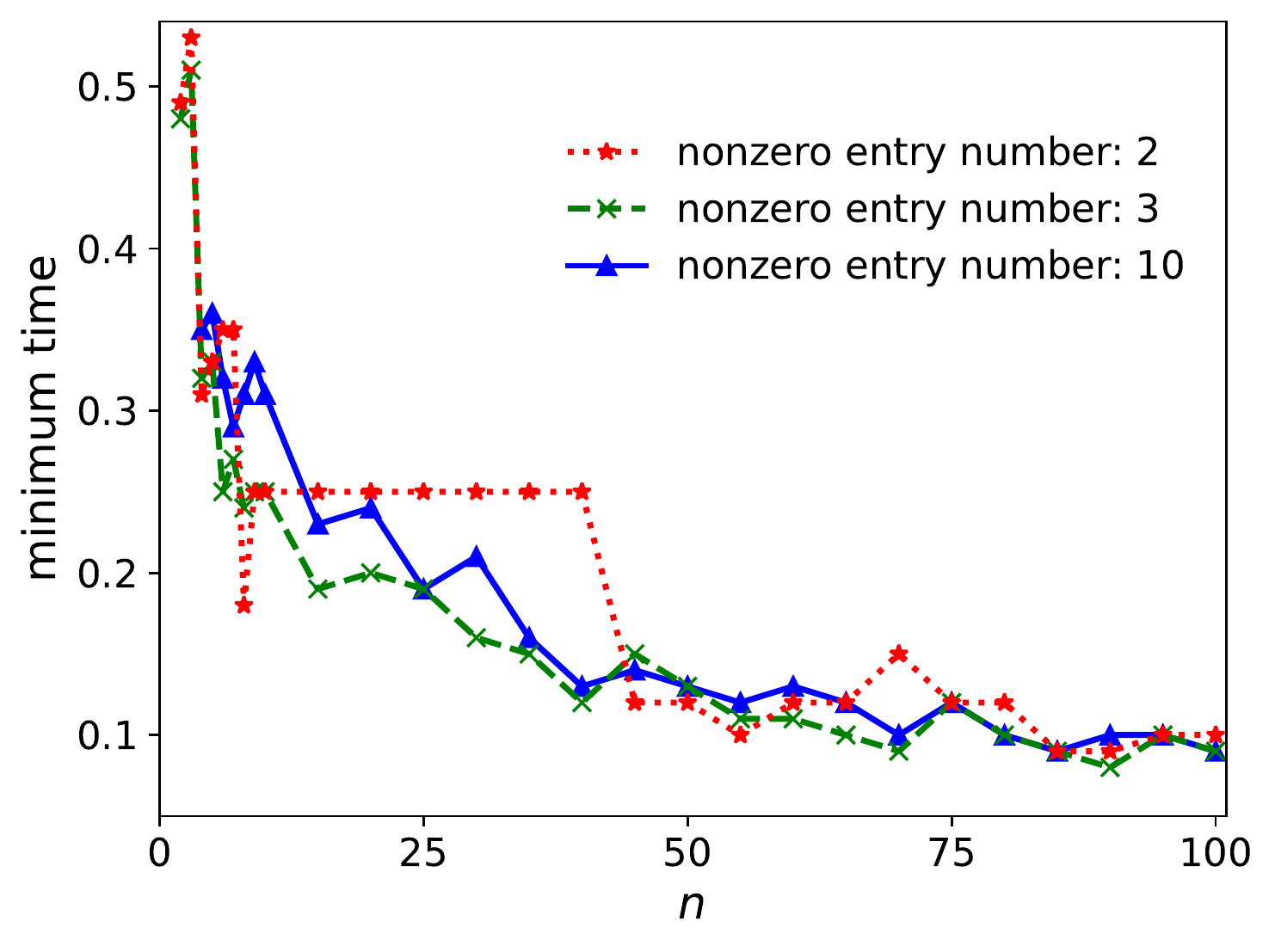}
\caption{The minimum time to reach the target as a function of spin number $n$ for $2000$
random states with $2$ nonzero entries (dotted-red-pentagram line), $3$ nonzero entries
(dashed-green-cross line), and $10$ nonzero entries (solid-blue-triangle line). The
parameters are set as $\Theta=\pi/2$, $B=0.5$, and $\omega/J=1$.}
\label{fig:n_mint}
\end{figure}
%========================================================================================

%==================================== Table =================================
\begin{table*}[tp]
\begin{tabular}{ccccccc}
\hline
\hline
\multirow{2}{*}{nonzero entry number~~} & \multicolumn{2}{c}{classification}
& \multicolumn{3}{c}{regression} & ~~calibration~~ \\
& ~~score~~ & ~~ratio~~ & ~~mean square error~~
& ~~$\tau_{\mathrm{learn}}$~~ & ~~true value~~ & ~~optimal time~~ \\
\hline
2 & 94.55$\%$ & 7.71$\%$ & 8.95$\times10^{-4}$ & 0.24 & 0.19 & 0.18 \\
3 & 89.67$\%$ & 5.85$\%$ & 2.06$\times10^{-2}$ & 0.18 & 0.25 & 0.24 \\
4 & 85.44$\%$ & 6.73$\%$ & 1.69$\times10^{-2}$ & 0.52 & 0.37 & 0.36 \\
5 & 81.52$\%$ & 9.48$\%$ & 1.54$\times10^{-2}$ & 0.54 & 0.24 & 0.24 \\
\hline
\hline
\end{tabular}
\caption{Results of CRC methodology in the categories of $2$, $3$, $4$,
and $5$ nonzero entries. }
\label{table:ising}
\end{table*}
%===========================================================================

%======================================== Figure ========================================
\begin{figure}[tp]
\centering\includegraphics[width=8.5cm]{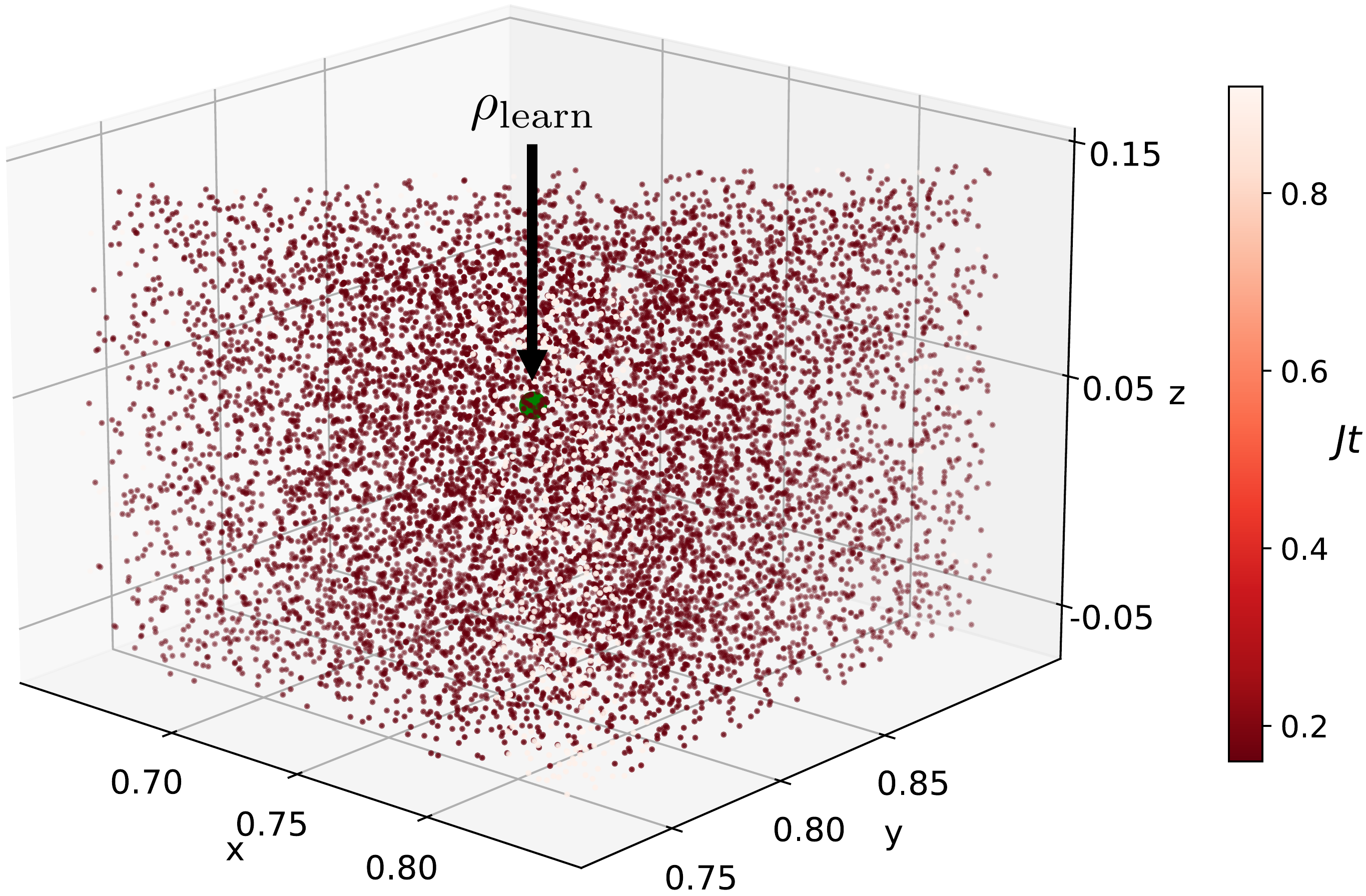}
\caption{Calibration result in the category of states with $2$ nonzero entries. The
green dot is the position of $\rho_{\mathrm{learn}}$. }
\label{fig:cali}
\end{figure}
%========================================================================================

In this section we show the OQSL in the case of the one-dimensional transverse Ising model, of which the Hamiltonian is
\begin{equation}
H/J=-\sum_{j=1}^n\sigma_j^z\sigma_{j+1}^z-g(t)\sum_{j=1}^n\sigma_j^x,
\end{equation}
where $J$ is the interaction strength between the qubits, and $g(t)$ is the time-dependent strength of the external field.
Here we consider that $g(t)=B\cos(\omega t)$.

Because of the enormous state space for this Hamiltonian, it is not easy to set up good training sets that are general enough
for the CRC methodology. To feasibly apply the CRC methodology, we need to analyze the state structure first and reduce the
state space for the study. A simple way to categorize the states is based on the number of nonzero entries in a certain basis.
Therefore, we analyze the evolution time to reach the target for the states with different numbers of nonzero entries in the
basis $\{\ket{\uparrow},\ket{\downarrow}\}^{\otimes n}$. Here $\ket{\uparrow}$ ($\ket{\downarrow}$) is the eigenvalue of
$\sigma_z$ with respect to the eigenvalue $1$ ($-1$). The evolution time to reach the target has been calculated for $2000$
random states in each category, and the minimum time is given in Fig.~\ref{fig:n_mint}. It can be seen that the minimum time
for the states with $2$ (dotted-red-pentagram line) and $3$ (dashed-green-cross line) nonzero entries is always lower than
that for the states with $10$ (solid-blue-triangle line) nonzero entries when $n$ is no larger than $100$. This phenomenon
indicates that in this example we only need to focus on the states with few nonzero entries for the study of OQSL.

In the meantime, we found an interesting phenomenon. The ratio of reachable states in the $2000$ random states basically
fits the function
\begin{equation}
\frac{1}{1+an^b e^{-cn^d}}.
\end{equation}
The parameters $a,b,c,d$ are $1.132$, $1.309$, $0.005$, $1.826$ for the category of $2$ nonzero entries, and $0.450$,
$1.654$, $0.034$, $1.355$ for the category of $3$ nonzero entries, and $0.613$, $0.842$, $0.007$, $1.754$ for the category
of $10$ nonzero entries. The fitting errors in three cases are 0.031, 0.028, and 0.031, respectively. For a large number 
of spins, basically all states can fulfill the target. We think a possible explanation for this phenomenon is that in a large Hilbert 
space, the number of target states for a given target is significantly large for any state, hence it is very easy to fulfill the target. 
The true ratio in this case and the physical mechanism behind it are still open questions and need to be further investigated in 
the future. 

Next, we perform the CRC methodology to evaluate the OQSL. Since we only need to focus on the states with few nonzero entries,
the CRC methodology is applied in the categories of states with $2$, $3$, $4$, and $5$ nonzero entries. The results are given
in Table~\ref{table:ising}. In all cases, $22500$ and $7500$ states and corresponding results ($0$ or $1$) consist of the
training and test sets in the classification process. The best score of the trained network we obtain is $94.55\%$, $89.67\%$,
$85.44\%$, and $81.52\%$ in the categories of $2$, $3$, $4$, and $5$ nonzero entries. The results show that $7.71\%$, $5.85\%$,
$6.73\%$, and $9.48\%$ states can fulfill the target in these categories. In the regression process, we also use $22500$ and
$7500$ states and the corresponding evolution time to reach the target as the training and test sets. The best mean square errors
of the trained network we obtain are $8.95\!\times\!10^{-4}$, $0.0206$, $0.0169$, and $0.0154$ in the categories of $2$, $3$,
$4$, and $5$ nonzero entries, and corresponding values of $\tau_{\mathrm{learn}}$ are $0.24$, $0.18$, $0.52$, and $0.54$. The
true values of the evolution time of $\rho_{\mathrm{learn}}$ are $0.19$, $0.25$, $0.37$, and $0.24$. The gap between
$\tau_{\mathrm{learn}}$ and the true values are majorly affected by the mean square errors, and it becomes difficult to obtain
a good mean square error with the increase of the nonzero entry number. About 10000 random states in the neighborhood of
$\rho_{\mathrm{learn}}$ are used in the process of calibration. These states share the same positions of nonzero entries with
$\rho_{\mathrm{learn}}$ and the differences of the norms and phases between them and $\rho_{\mathrm{learn}}$ are less than
$0.1$. The calibration in the category of $2$ nonzero entries is shown in Fig.~\ref{fig:cali}. The $x$ and $y$ axes are the
norms of the nonzero entries and the $z$ axis is the phase difference between these two entries. The green dot is the position
of $\rho_{\mathrm{learn}}$. The other three categories are not shown since the parameters are larger than $3$. After the
calibration, the optimal evolution time in these categories is $0.18$, $0.24$, $0.36$, and $0.24$. Hence, the final
evaluation of OQSL in this example is $0.18$, which can be realized by certain states with $2$ nonzero entries. 

In this example, the time cost for the generation of training sets for one category is about one day on a work station with $12$  
threads, and those for the training of the neural networks in the classification and regression processes are only several minutes. 
Moreover, the time cost of the calibration for one category is about several hours. Hence, for large-scale systems the major time 
cost to implement the CRC methodology is the generation of training sets.

%\end{document}


\begin{thebibliography}{99}
\bibitem{footnote}
The actual computing time is less than the evaluation since parallel computing is applied.
\end{thebibliography}
\end{document}